\documentclass[fleqn,usenatbib]{mnras}

\usepackage{newtxtext,newtxmath}
\usepackage[T1]{fontenc}
\usepackage{ae,aecompl}

\usepackage{graphicx}                                      % Including figure files
\usepackage{threeparttable}                                % for last table notes
\usepackage{natbib}

\title[Variable dust emission by WR stars]{\bf Variable dust emission by 
WC type Wolf-Rayet stars observed in the NEOWISE-R survey}
\author[P. M. Williams] {P. M. Williams\thanks{E-mail: pmw@roe.ac.uk}\\
Institute for Astronomy, University of Edinburgh, Royal Observatory, Edinburgh EH9 3HJ\\}

\date{Accepted 2019 June  .
      Received 2019 June 26;
      in original form 2019 May 17}
\pubyear{2019}

\begin{document}

\pagerange{\pageref{firstpage}--\pageref{lastpage}}

\label{firstpage}
\maketitle

\begin{abstract}
Photometry at 3.4 and 4.6 $\micron$ of 128 Population~I WC type Wolf-Rayet 
stars in the Galaxy and 12 in the Large Magellanic Cloud (LMC) observed in the 
{\em WISE} NEOWISE-R survey was searched for evidence of circumstellar 
dust emission and its variation. 
Infrared spectral energy distributions (SEDs) were assembled, making use of 
archival $r$, $i$, $Z$ and $Y$ photometry to determine reddening and stellar 
wind levels for the WC stars found in recent IR surveys and lacking optical 
photometry. 
From their SEDs, ten apparently non-variable stars were newly identified as 
dust makers, including three, WR\,102-22, WR\,110-10 and WR\,124-10, having 
subtype earlier than WC8--9, the first such stars to show this phenomenon. 
The 11 stars found to show variable dust emission include six new episodic 
dust-makers, WR\,47c, WR\,75-11, WR\,91-1, WR\,122-14 and WR\,125-1 in the 
Galaxy and HD 38030 in the LMC. 
Of previously known dust makers, NEOWISE-R photometry of WR\,19 captured its 
rise to maximum in 2018 confirming the 10.1-y period, that of WR\,125 the 
beginning of a new episode of dust formation suggesting a period near 28.3~y.\ 
while that of HD~36402 covered almost a whole period and forced revision of 
it to 5.1~y.
\end{abstract}


\section{Introduction}

Wolf-Rayet (WR) stars are in an advanced stage of evolution, losing mass 
through dense stellar winds which give rise to their characteristic 
emission-line spectra. 
One of the earliest results from infrared (IR) astronomy was the discovery 
of `excess' IR radiation by heated circumstellar dust from a variety of 
stars having emission lines. Amongst these were four WC9 type 
Wolf--Rayet (WR) stars observed by \citet*{ASH} and, at longer wavelengths, 
by \citet{GH74}. \citet*{CBK} measured optical--IR energy distributions of a 
sample of WR stars and showed that they could be matched by either free-free 
or graphite dust emission. Early observations \citep{HGSS} also showed 
variations in the IR emission by two out of a sample of ten WR stars, 
HD~193793 (WR\,140) and HD~192641 (WR\,137), which the authors interpreted 
in terms of changes in the electron densities, radii and mass-loss rates 
affecting their free-free emission. These variations had to be reinterpreted 
as fading circumstellar dust emission when the subsequent brightening of 
these stars in the IR showed spectral energy distributions (SEDs) 
characteristic of heated dust emission \citep[respectively]{WBLSA,W137}. 
Such stars can be described as episodic dust makers to distinguish them 
from the (apparently) constant dust makers like the WC9 stars found in 
the early studies referred to above.

Continued IR photometry of WR\,140 discovered another dust maximum in 1985, 
leading to a period of 7.9~y and allowing re-interpretation of the previously 
known radial velocity (RV) variations in terms of a very eccentic orbit having 
periastron passage and closest approach of the WC7 and O5 stars coinciding 
with maximum dust emission \citep{W140}. 
Variations in its strong radio and X-ray emission were also tied to its binary 
orbit, so that WR\,140 has become the prototype colliding-wind binary (CWB). 
It has continued to attract investigations across the spectrum from radio 
to X-ray (e.g. \citet{VLBA140}, \citet{SUZAKU140} and references therein) and 
continued refinement of its orbit from both RV and astrometric studies 
\citep{Fahed140,Monnier140}. 
\citet{Usov} showed that dust could form in the collisionally shocked WC7 
wind if it cooled efficiently. This was the first step in tackling the 
underlying problem presented by dust formation by some WC type WR stars: 
the great difficulty of forming dust in such hostile environments 
\citep*{HGG,WHT}. 
High-density structures in the WC winds are required to allow dust formation 
and these can be provided by shocks if their winds collide with those of 
massive companions. The winds in CWBs collide all the 
time but dust formation is rarer and requires particular conditions, which are 
satisfied in WR\,140 for only a brief time during periastron passage in its 
very elliptical ($e$ = 0.896) orbit \citep{Fahed140}, presumably owing to 
higher pre-shock wind density when the stars are closest \citep{WIAUC169}. 
Comparison of the physical conditions in wind-collision regions of
well-studied CWBs such as WR\,140 at phases when dust does and does not form 
may provide the key to understanding dust formation in WR systems.

Spurred by the episodic formation of dust by WR\,140 when the stars reached 
a critical separation in their orbit,  \citet{WASP22} suggested that the 
persistent WC8--9 dust makers might also be binaries,  but in circular 
orbits having stellar separations and pre-shock densities always conducive 
to dust formation -- as beautifully demonstated by the rotating dust 
`pinwheel' made by the prototypical WC9 dust maker WR\,104 \citep{Tuthill104} 
observed in high resolution near-IR images. 
From their `pinwheel images of WR\,104, \citet{Tuthill104b} showed that the 
dust was being made by the WC9 and companion stars moving in a circular orbit 
and that the IR flux level, and hence the dust formation rate, did not vary 
with orbital phase.
The circular orbit results in constant separation of the stars and constant 
densities in the stellar winds before they collide, evidently accounting 
for the constant dust formation making the pinwheel. The corollary of 
this is that, within the paradigm of WR dust formation in CWBs, constancy 
of dust formation implies that such binaries have circular orbits.

Long-term near-IR photometric histories of 14 other WC8--9 dust-makers 
compiled by \citet{WPotsdam} showed no significant ($\sigma > 0.04$) variation 
for most of them, with only two to be variable, WR\,65 and WR\,112. 
If most dust-making WC8--9 stars turn out to be binaries having circular 
orbits, this suggests that they may have suffered interaction and 
circularisation in the course of their evolution (cf. \citet{Tuthill104b} 
regarding WR\,104), perhaps to a greater extent than WC stars of earlier 
subtypes, amongst which dust formation is much rarer and episodic. 

The goal of the present investigation is to search for variation in the IR 
emission by WR circumstellar dust in a large and homogeneous data set. 
The Near-Earth Object {\em WISE} Reactivation (NEOWISE-R) mission 
\citep{NEOWISER} is well suited for this investigation, providing 
observations over five years and allowing characterisation of the 
variability such as periodic variation which could indicate CWB-modulated 
dust formation, or slow variations similar to those shown by the longer 
period dust makers WR\,137 and WR\,140. 
A by-product is the identification of more dust makers amongst the recently 
catalogued WR stars. 

%------------------------------------------------------------------------------
\section{Data sets and search for variability}    % ---------------------------
\label{SData}
%------------------------------------------------------------------------------

\begin{table*} 
%\centering 
\caption{Galactic program stars with mean NEOWISE-R $W1$, its dispersion $\sigma$ 
(rounded to 0.01 mag.) and the long-term variability metric, $R$, defined in the 
text, followed by the corresponding quantities for $W2$. 
An asterisk in the `N' column denotes a star discussed individually in Section 
\ref{Sindividual}, while `d' denotes other stars found or confirmed to be dust 
makers in this study and `w' the possible dust makers found here to have stellar-wind SEDs.} 
\label{TSum}
\begin{tabular}{cllrrrrrrc}
\hline
   WR   &    Name          & Spectrum &  $W1$ & $\sigma$ & $R$  &  $W2$ & $\sigma$ & $R$  & N \\ 
\hline  
  4     & HD 16523         & WC5+?    &  7.67 & 0.01     & 0.4  &  7.35 & 0.01     &  0.4 &   \\  
  5     & HD 17638         & WC6      &  7.24 & 0.02     & 0.3  &  7.01 & 0.01     &  0.3 &   \\
  9     & HD 63099         & WC5+O7   &  6.97 & 0.08     & 1.0  &  6.79 & 0.08     &  2.4 &   \\  
 13     & Ve6-15           & WC6      &  8.56 & 0.01     & 0.3  &  8.16 & 0.01     &  0.3 &   \\
 17     & HD 88500         & WC5      &  8.84 & 0.01     & 0.3  &  8.52 & 0.01     &  0.2 &   \\
 19     & LS 3             & WC4pd+O9 &  7.74 & 1.05     & 18.8 &  7.28 & 1.32     & 16.6 & * \\
 23	    & HD 92809         & WC6      &  6.28 & 0.04     & 0.2  &  6.28 & 0.04     &  0.4 &   \\
 27	    & MS 1             & WC6+a    &  7.91 & 0.01     & 0.5  &  7.50 & 0.01     &  0.4 &   \\
 30     & HD 94305         & WC6+O6-8 &  8.96 & 0.03     & 1.2  &  8.62 & 0.03     &  1.5 &   \\
 31c    & SMSP 4           & WC6      &  8.95 & 0.01     & 0.3  &  8.58 & 0.02     &  0.5 &   \\
 33	    & HD 95345         & WC6      &  9.56 & 0.01     & 0.3  &  9.22 & 0.01     &  0.4 &   \\
 41	    & LS 7             & WC5+OB?  &  9.81 & 0.01     & 0.2  &  9.44 & 0.01     &  0.2 &   \\
 42	    & HD 97152         & WC7+O7V  &  6.43 & 0.06     & 0.3  &  6.45 & 0.02     &  0.4 &   \\	
 44-1   & SMG09 740\_16    & WCE      & 10.08 & 0.00     & 0.2  &  9.74 & 0.01     &  0.4 &   \\
 45	    & LSS 2423         & WC6      &  8.86 & 0.01     & 0.4  &  8.47 & 0.01     &  0.4 &   \\
 46-7   & MV09             & WC5-7    &  8.83 & 0.35     & 12.7 &  8.16 & 0.34     & 14.1 & * \\
46-10   & SMG09 791\_12c   & WCE      & 11.01 & 0.02     & 0.3  & 10.60 & 0.01     &  0.3 &   \\
46-11   & SMG09 808\_14    & WCE      & 10.40 & 0.03     & 1.1  & 10.01 & 0.03     &  1.2 &   \\
46-13   & SMG09 807\_13    & WC7      & 10.54 & 0.01     & 0.3  & 10.10 & 0.01     &  0.5 &   \\
46-18   & RC17 E3          & WC6-7    &  9.79 & 0.00     & 0.2  &  9.22 & 0.01     &  0.3 &   \\  
 47c    & SMSNPL 7         & WC5      &  9.42 & 0.12     & 3.7  &  9.10 & 0.14     &  4.9 & * \\
 47-2   & SMG09 832\_25    & WC5-6    &  9.63 & 0.01     & 0.5  &  9.26 & 0.01     &  0.3 &   \\ 
 47-3   & SMG09 856\_13c   & WC5-6    & 10.42 & 0.01     & 0.5  & 10.07 & 0.01     &  0.5 &   \\
 48-1   & HDM 5	           & WC7      &  8.72 & 0.01     & 0.3  &  8.26 & 0.01     &  0.4 &   \\ 
 48b    & SMSNPL 8         & WC9d     &  7.13 & 0.11     & 0.8  &  6.59 & 0.08     &  1.0 &   \\
48-2    & Danks 2-3	       & WC7-8    &  8.06 & 0.03     & 0.5  &  7.66 & 0.03     &  0.4 &   \\ 
48-3    & SMG09 845\_34    & WC8      &  8.20 & 0.02	 & 0.5  &  7.95 & 0.01     &  0.4 &   \\ % Svet X-ray
 50	    & Th2 84           & WC7+OB	  &  8.52 & 0.00	 & 0.2  &  8.22 & 0.01     &  0.5 &   \\
 52	    & HD 115473        & WC4      &  7.28 & 0.02     & 0.2  &  7.03 & 0.01     &  0.5 &   \\
 57     & HD 119078        & WC8      &  7.16 & 0.04     & 0.5  &  6.82 & 0.05     &  1.4 &   \\
 59-1	& SMG09 883\_18    & WCE      & 10.13 & 0.01     & 0.3  &  9.59 & 0.01     &  0.4 &   \\
 59-2   & SMG09 885\_11    & WC5-6	  & 10.32 & 0.03     & 0.5  &  9.79 & 0.02     &  0.4 &   \\ % wind
 60	    & HD 121194	       & WC8      &  6.90 & 0.03     & 0.3  &  6.70 & 0.03     &  0.6 &   \\
 60-1   & Sm09 897.5       & WC8      &  9.20 & 0.03     & 1.3  &  8.42 & 0.03     &  1.1 &   \\ % wind
 60-2   & SMG09 903\_15c   & WC8      &  8.16 & 0.03     & 1.7  &  7.39 & 0.03     &  1.6 & d \\ % dust
 60-3   & MDM11-11         & WC7      & 10.80 & 0.03     & 0.4  & 10.31 & 0.03     &  0.4 & * \\
 60-4   & MDM11-12         & WC8      &  9.63 & 0.11     & 5.2  &  8.79 & 0.14     &  6.5 & * \\
 60-5   & WR 60a R11b      & WC7      &  9.00 & 0.00     & 0.3  &  8.69 & 0.01     &  0.5 &   \\
 60-7   & RC17 B51         & WC7-8    &  9.30 & 0.04     & 2.0  &  8.71 & 0.04     &  1.5 &   \\  
 61-3   & MDM11-13         & WC9      &  9.32 & 0.03	 & 1.1  &  8.68 & 0.03     &  1.4 &   \\
 64	    & BS 3             & WC7      & 10.71 & 0.03	 & 0.6  & 10.41 & 0.02     &  0.6 &   \\
 67-2   & WR67b R11a       & WC7      &  7.88 & 0.01     & 0.3  &  7.49 & 0.01     &  0.3 &   \\
 68	    & BS 4             & WC7      &  8.31 & 0.01	 & 0.3  &  7.93 & 0.01     &  0.4 &   \\
 70-3   & SMG09 1011\_24   & WC7      &  8.49 & 0.01	 & 0.4  &  8.14 & 0.01     &  0.4 &   \\
 70-9   & MDM11 17         & WC8      & 10.22 & 0.01	 & 0.3  &  9.55 & 0.01     &  0.5 &   \\
 70-12  & SFZ12 1038-22L   & WC7:     &  8.58 & 0.02	 & 0.4  &  8.03 & 0.00     &  0.2 &   \\
 70-13  & RC17 B105        & WC8d     &  8.61 & 0.05     & 1.6  &  7.92 & 0.06     &  2.6 &   \\ 
 72-1   & HDM6             & WC9      &  7.91 & 0.01     & 0.4  &  7.45 & 0.01     &  0.3 &   \\
 72-2   & SMG09 1053\_27   & WC8      &  7.24 & 0.03     & 0.3  &  6.91 & 0.01     &  0.3 &   \\ 
 72-3   & MDM11 18         & WC9d?    & 10.49 & 0.02	 & 0.6  &  9.79 & 0.02     &  0.7 & w \\ % wind 
 72-4   & SFZ12 1051-67L   & WC7:     &  9.94 & 0.03	 & 0.4  &  9.45 & 0.02     &  0.3 &   \\
 73-1   & SMG09 1059\_34   & WC7      & 10.58 & 0.05	 & 0.8  & 10.13 & 0.03     &  0.8 &   \\
 74-3   & SFZ12 1077-55L   & WC6:     & 10.74 & 0.02	 & 0.3  & 10.25 & 0.02     &  0.6 &   \\ 
 75a    & SMSNPL 15	       & WC9      &  7.91 & 0.01     & 0.4  &  7.50 & 0.01     &  0.6 &   \\
 75aa   & HBD 1	           & WC9d     &  8.17 & 0.06     & 2.7  &  7.51 & 0.06     &  2.7 & * \\
 75b    & SMSNPL 16	       & WC9      &  7.54 & 0.03     & 0.4  &  7.25 & 0.01     &  0.4 &   \\
 75c    & HBD 2	           & WC9      & 10.02 & 0.02     & 1.1  &  9.58 & 0.03     &  1.6 &   \\
 75d    & HBD 3	           & WC9      &  8.21 & 0.08     & 1.6  &  7.65 & 0.09     &  1.8 & * \\  % dust
 75-1   & SMG09 1081\_21   & WC8      &  9.76 & 0.03     & 0.6  &  9.40 & 0.02     &  0.9 &   \\
 75-2   & SMG09 1093\_34   & WC8      & 10.22 & 0.02     & 0.2  &  9.60 & 0.03     &  0.6 &   \\ 
\hline
\end{tabular}                 % 60 stars so far, WR 70-10 removed: 3" neighbour
\end{table*}

\begin{table*}
%\centering
\contcaption{Galactic program stars with mean NEOWISE-R $W1$ and $W2$ and their dispersions.}
\begin{tabular}{cllrrrrrrc}
\hline
   WR   &    Name          & Spectrum  &  $W1$ & $\sigma$ & $R$  &  $W2$ & $\sigma$ &  $R$ & N \\ 
\hline
 75-3   & SMG09 1093\_33   & WC8       & 10.36 & 0.02     & 0.4  &  9.79 & 0.01     &  0.3 &   \\
 75-5   & SMG09 1096\_22   & WC8       & 10.46 & 0.03     & 0.5  &  9.89 & 0.01     &  0.4 &   \\
 75-7   & MDM11 22         & WC9       & 10.59 & 0.03     & 0.4  &  9.95 & 0.02     &  0.4 & d \\ 
 75-11  & MDM11-26         & WC9d?     &  9.43 & 0.06     & 1.3  &  8.90 & 0.07     &  2.2 & * \\ 
 75-14  & SFZ12 1085-72L   & WC9       & 10.14 & 0.05     & 0.6  &  9.70 & 0.04     &  0.7 &   \\
 75-15  & SFZ12 1085-69L   & WC8       &  9.95 & 0.03     & 0.7  &  9.24 & 0.03     &  0.8 & d \\
 75-16  & SFZ12 1085-83L   & WC8       & 10.55 & 0.03     & 0.4  &  9.92 & 0.02     &  0.5 &   \\
 75-19  & SFZ12 1093-140L  & WC7:      & 11.31 & 0.03     & 0.4  & 10.80 & 0.06     &  1.1 &   \\
 75-23  & SFZ12 1106-31L   & WC9       &  8.34 & 0.04     & 1.3  &  7.88 & 0.05     &  2.2 &   \\
 75-24  & SFZ12 1105-76L   & WC8       & 10.49 & 0.01     & 0.2  &  9.94 & 0.01     &  0.3 &   \\
 76-10  & SFZ12 1109-74L   & WC7:      &  9.85 & 0.04     & 0.7  &  9.30 & 0.04     &  0.9 &   \\
 77     & He3-1239         & WC8+O8	   &  7.45 & 0.05     & 0.9  &  7.04 & 0.03     &  1.0 &   \\
 77t    & HBD 5	           & WC9d      &  6.89 & 0.18     & 2.0  &  6.36 & 0.15     &  2.1 & * \\
 81     & He3-1316         & WC9       &  6.06 & 0.06     & 0.4  &  5.90 & 0.02     &  0.2 &   \\
 82-2   & KSF14 1178-66B   & WC9       &  9.48 & 0.03     & 0.7  &  9.04 & 0.03     &  0.9 &   \\
 84-2   & SFZ12 1181-82L   & WC8       &  9.80 & 0.03     & 0.3  &  9.31 & 0.02     &  0.4 &   \\
 84-5   & SFZ12 1189-110L  & WC9       & 10.76 & 0.05     & 1.0  & 10.24 & 0.07     &  2.2 &   \\
 91-1   & SMG09 1222\_15   & WC7       &  9.39 & 0.25     & 6.9  &  8.31 & 0.37     & 14.5 & * \\
 92     & HD 157451        & WC9       &  8.23 & 0.03     & 1.4  &  7.82 & 0.04     &  1.5 &   \\
 94-1   & SFZ12 1245-23L   & WC9       &  9.59 & 0.03     & 0.7  &  8.63 & 0.03     &  0.7 &   \\ 
 98-1   & SFZ12 1269-166L  & WC8       &  9.44 & 0.02     & 0.5  &  8.92 & 0.02     &  0.4 &   \\
101     & DA 3             & WC8       &  6.84 & 0.06     & 0.4  &  6.60 & 0.02     &  0.4 &   \\
102-22  & WR 1327-14AF     & WC7       &  9.98 & 0.05     & 0.9  &  9.54 & 0.08     &  2.0 & d \\
111-3   & SMG09 1385\_24   & WC8       &  7.60 & 0.04     & 0.8  &  7.10 & 0.03     &  1.1 &   \\
111-7   & SFZ12 1395-86L   & WC8       & 10.58 & 0.03	  & 0.4  &  9.75 & 0.02     &  0.4 &   \\ 
111-10  & KSF14 1389-4AB6  & WC7       & 10.29 & 0.02	  & 0.4  &  9.47 & 0.02     &  0.6 & d \\
113-2   & SMG09 1425\_47   & WC5-6     &  7.81 & 0.01     & 0.3  &  7.35 & 0.01     &  0.4 &   \\
118-4   & MDM11 39         & WC8       &  8.64 & 0.01	  & 0.3  &  8.06 & 0.01     &  0.3 &   \\
118-8   & SFZ12 1487-80L   & WC9       &  9.94 & 0.03	  & 0.6  &  9.11 & 0.03     &  1.0 &   \\
119-2   & MDM11 42         & WC8       &  8.81 & 0.02	  & 0.4  &  8.41 & 0.04     &  0.6 &   \\
119-4   & KSF14 1495-1D8A  & WC8-9     & 10.12 & 0.08     & 0.5  &  9.44 & 0.02     &  0.3 &   \\ 
120-1   & HDM 13           & WC9       &  8.75 & 0.02     & 0.4  &  8.26 & 0.01     &  0.4 &   \\
120-5   & SCB12 2w02       & WC8       &  8.72 & 0.04     & 1.2  &  8.25 & 0.05     &  1.7 &   \\
120-11  & SFZ12 1495-32L   & WC8       &  9.58 & 0.02     & 0.6  &  9.11 & 0.02     &  0.6 &   \\
120-13  & SFZ12 1522-55L   & WC9       & 10.27 & 0.04     & 0.9  &  9.71 & 0.05     &  2.1 &   \\
120-14  & SCB12 2w03       & WC8       & 10.67 & 0.01     & 0.3  & 10.22 & 0.01     &  0.4 &   \\
120-15  & SCB12 2w04       & WC8       &  9.57 & 0.02     & 0.3  &  9.50 & 0.02     &  0.6 &   \\ 
120-16  & KSF14 1514-AA0   & WC8       & 10.12 & 0.01     & 0.2  & 10.19 & 0.01     &  0.3 &   \\ 
120-17  & KSF14 1509-2E64  & WC9       &  9.75 & 0.06     & 0.5  &  9.01 & 0.07     &  0.8 & d \\
121-4   & MDM11 49         & WC8       &  9.79 & 0.02     & 0.2  &  9.19 & 0.03     &  1.3 &   \\
121-5   & SCB12 2w07       & WC8       &  9.00 & 0.01     & 0.3  &  8.13 & 0.01     &  0.3 &   \\
121-10  & SCB12 2w10       & WC8       & 10.82 & 0.05     & 0.9  & 10.09 & 0.05     &  1.5 &   \\
121-13  & KSF14 1541-187C  & WC8       & 10.06 & 0.05     & 0.7  &  9.30 & 0.05     &  1.1 &   \\
122-1   & IPHAS J190015.86+000517.3 & WC8  &  9.38 & 0.03 & 1.5  &  8.83 & 0.04     &  1.8 & d \\ 
122-7   & SFZ12 1563-66L   & WC8       &  9.94 & 0.03     & 0.5  &  9.19 & 0.02     &  0.6 &   \\
122-8   & SFZ12 1563-89L   & WC7:      & 11.40 & 0.04     & 0.4  & 10.70 & 0.02     &  0.5 &   \\ 
122-9   & SFZ12 1567-51L   & WC7:      &  9.73 & 0.01     & 0.2  &  9.22 & 0.01     &  0.5 &   \\ 
122-14  & KSF14 1553-15DF  & WC8       &  9.76 & 0.10     & 3.3  &  8.69 & 0.11     &  4.2 & * \\
123-4   & SFZ12 1603-75L   & WC8       & 10.65 & 0.02     & 0.6  &  9.89 & 0.03     &  1.2 &   \\
123-5   & SCB12 2w11       & WC7       & 10.50 & 0.01     & 0.2  &  9.82 & 0.01     &  0.4 &   \\
124-2   & SMG09 1671\_5    & WC8       &  9.96 & 0.02     & 0.2  &  9.55 & 0.01     &  0.2 &   \\
124-3   & MDM11 56         & WC7       & 10.18 & 0.01     & 0.4  &  9.74 & 0.01     &  0.4 &   \\
124-5   & MDM11 58         & WC8-9d?   &  9.52 & 0.02     & 0.7  &  8.86 & 0.03     &  0.9 & w \\ % wind
124-6   & MDM11 59         & WC7       &  8.51 & 0.01     & 0.2  &  7.92 & 0.01     &  0.3 &   \\
124-7   & MDM11 60         & WC8d      &  8.39 & 0.04     & 0.5  &  7.75 & 0.09     &  0.8 &   \\
124-9   & SFZ12 1670-57L   & WC6:      & 11.11 & 0.03     & 0.4  & 10.60 & 0.01     &  0.2 &   \\
124-10  & SFZ12 1669-24L   & WC6       & 10.49 & 0.02     & 0.6  &  9.95 & 0.02     &  0.8 & d \\
124-16  & KSF14 1647-1E70  & WC8:      & 10.91 & 0.04     & 0.9  &  9.80 & 0.03     &  1.3 &   \\
124-19  & KSF14 1660-1169  & WC6:      & 11.61 & 0.01     & 0.2  & 11.15 & 0.02     &  0.4 &   \\
124-20  & KSF14 1697-38F   & WC9       &  8.53 & 0.05     & 1.9  &  7.88 & 0.05     &  2.3 & d \\
124-22  & KSF14 1695-2B7   & WC9       &  8.40 & 0.04     & 1.9  &  7.71 & 0.04     &  1.7 & d \\
125     & IC 14-36         & WC7ed+O9  &  7.73 & 0.07     & 2.0  &  7.38 & 0.12     &  6.2 & * \\
125-1   & HDM 15           & WC8       &  8.53 & 0.04     & 2.1  &  8.09 & 0.07     &  3.2 & * \\	
\hline
\end{tabular}             
\end{table*}

\begin{table*}
%\centering
\contcaption{Galactic Program stars with mean NEOWISE-R $W1$ and $W2$ and their dispersions.}
\begin{tabular}{cllrrrrrrc}
\hline
   WR   &    Name         & Spectrum  &  $W1$ & $\sigma$ & $R$ & $W2$ & $\sigma$ & $R$  & N \\ 
\hline  
132	    & HD 190002  	  & WC6+?     &  8.75 & 0.01     & 0.4 &  8.38 & 0.01    &  0.3 &   \\
143     & HD 195177       & WC4+Be    &  6.87 & 0.23     & 2.2 &  6.75 & 0.17    &  4.4 &   \\
144	    & MHM 19-1        & WC4       &  7.18 & 0.01     & 0.2 &  6.86 & 0.01    &  0.2 &   \\
150	    & ST 5            & WC5       &  9.37 & 0.00	 & 0.3 &  8.99 & 0.01    &  0.4 &   \\
154	    & HD 213049	      & WC6       &  8.01 & 0.01     & 0.4 &  7.65 & 0.01    &  0.4 &   \\
\hline
\end{tabular} 
\end{table*}

\begin{table*}
%\centering
\caption{Large Magellanic Cloud program stars with mean 
NEOWISE-R $W1$, its dispersion $\sigma$ and $R$, followed by the same quantities for $W2$.}
\label{TLMC}
\begin{tabular}{cllrrrrrrc}
\hline
 BAT99  & HD/Brey      & Spectrum  &  $W1$ & $\sigma$ &  $R$ & $W2$  & $\sigma$ &  $R$ & N \\
\hline
   8    & 32257  & WC4             & 13.49 & 0.03     &  0.3 & 13.18 &   0.02   &  0.2 &   \\
   9    & 32125  & WC4             & 13.52 & 0.03     &  0.6 & 13.22 &   0.01   &  0.1 &   \\
  11    & 32402  & WC4             & 12.38 & 0.01     &  0.2 & 12.12 &   0.01   &  0.2 &   \\
  34    & 36156  & WC4+OB          & 12.81 & 0.02     &  0.4 & 12.56 &   0.02   &  0.4 &   \\
  38    & 36402  & WC4(+O?)+O8I:   &  9.75 & 0.34     & 10.5 &  8.89 &   0.34   &  8.8 & * \\
  39    & 36521  & WC4+O           & 12.59 & 0.01     &  0.3 & 12.53 &   0.03   &  0.3 &   \\
  52    & 37026  & WC4             & 13.04 & 0.01     &  0.2 & 12.70 &   0.01   &  0.2 &   \\
  53    & 37248  & WC4+O9          & 13.01 & 0.02     &  0.3 & 12.85 &   0.02   &  0.3 &   \\
  61    & 37680  & WC4             & 12.48 & 0.01     &  0.2 & 12.16 &   0.01   &  0.2 &   \\
  84    & 38030  & WC4             & 12.70 & 0.42     &  8.8 & 12.21 &   0.69   & 10.2 & * \\
  87    & Br 70  & WC4+OB?         & 13.52 & 0.02     &  0.2 & 13.30 &   0.02   &  0.2 &   \\
 125    & 38448	 & WC5+O7          & 12.24 & 0.05     &  1.0 & 11.98 &   0.06   &  1.0 &   \\ 
\hline
\end{tabular}
\end{table*}

The principal data set on which this study is based is the 2019 data Release 
of the NEOWISE-R survey.
The wavelengths of the $W1$ and $W2$ bands, 3.4 and 4.6$\micron$, are well 
placed for observing $T_g \sim$ 1000-K circumstellar dust emission.
Synthetic $W1$ and $W2$ magnitudes calculated from a model stellar wind and 
heated carbon dust show that the $W1$--$W2$ colour is a good measure of the 
average dust temperature but not of the amount of dust if $T_g >$ 1000~K 
because its $W1$--$W2$ is then similar to that of the wind.
The instrument has the sensitivity to cover the fainter and heavily reddened 
WR stars found in recent IR surveys, e.g. \citet*{MV09}, \citet{SMG09} but, 
on the other hand, most of the WR dust emitters found in the earlier studies 
are too bright for NEOWISE-R (see below), so there is little overlap with the 
earlier studies. The only exceptions are the episodic dust makers WR\,19 and 
WR\,125, whose near-IR fluxes had faded to wind level before the first 
NEOWISE-R observations, and the variable dust maker HD 36402 \citep{W36402} 
in the Large Magellanic Cloud (LMC).

The NEOWISE-R data were collected in around ten `visits', akin to observing 
runs, each including 12 or more observations taken at intervals of one or more 
94-min orbital passages and spread over several days, with the visits separated 
by about six months as the Sun-synchronous orbit of the satellite followed the 
Earth in its orbit. The length of visit and number of observations in each 
depend on the overlap of the survey strips, which increase with increasing 
ecliptic latitude as the overlap increases. The observations were taken 
between late 2013 and late 2018. The cadence of the observations makes 
them most useful for studying variations over short ($\sim$ 1--2~d.) and long 
($\sim$ 1--5~y.) time-scales, of which the latter are of interest for the 
present investigation.

NEOWISE-R followed the original {\em Wide-field Infrared Explorer (WISE)} 
mission \citep{WISE}, which surveyed the whole sky (All-Sky survey) in four 
bands: $W1$, $W2$, $W3$ and $W4$ at 3.4, 4.6, 11 and 22 $\micron$ respectively. 
As the cryogen became exhausted, the $W4$ observations were dropped and 
surveying continued in three bands (3-Band Cryo) for almost two months, 
after which $W3$ observations also ceased and surveying continued in 
$W1$ and $W2$ in the Post-Cryo {\em NEOWISE} survey \citep{NEOWISE}.
Altogether, these surveys provide at least another two sets of $W1$ and 
$W2$ photometry, observed mostly in 2010.
There is a significant difference, however, between the NEOWISE-R data and 
those from the All-Sky and 3-Band Cryo surveys: the profile-fit brightnesses
of sources brighter than the saturation limits $W1 \simeq 8$ and 
$W2 \simeq 7$ are significantly overestimated in the NEOWISE-R data. 
The offsets \citep[figure 6]{NEOWISER} rise to 0.8--0.9 mag. with considerable 
dispersion. Such stars were not initially excluded from the present study: 
data were extracted where possible, and the offsets borne in mind during 
their interpretation. Useful data were recovered for those as bright as 
$W1$ and $W2 \sim 6$ in NEOWISE-R, corresponding to $W1 \sim 6.5$ 
and $W2 \sim 6.2$ in the cryogenic (All-Sky and 3-Band Cryo) surveys, with 
slightly more scatter in the individual magnitudes.
 
The NEOWISE-R Single Exposure Source Database in the NASA/IPAC Infrared 
Science Archive (IRSA) data was searched at the positions of all WC stars 
in V~1.22\footnote{{\tt http://pacrowther.staff.shef.ac.uk/WRcat/index.php},
dated Mar 2019}
of the Galactic Wolf Rayet Catalogue \citep{RC15}. Besides saturation 
of the brighter sources, the principal limitation was source confusion, 
given the 6.1 and 6.4 arc sec psf $W1$ and $W2$ beam sizes \citep{WISE}. 
This effectively excluded the WR stars in crowded regions such as near 
the direction of the Galactic Centre or massive star clusters, 
as well as field WR stars which happened to have close neighbours. 
The Galactic programme stars are listed in Table \ref{TSum}. The WR numbers 
are from the Galactic Wolf-Rayet Catalogue, together with earlier or  
`discovery' names. 
The spectral types, including `d' to mark dust makers, are from the Catalogue, 
as updated by \citet{RC15} where appropriate. Stars found to be dust makers in 
the present study are flagged `d' in the `N' column while those catalogued as 
possible dust makers (`d?') are flagged `d' or `w' where their SEDs were found 
to show dust or just stellar wind emission.

In addition to the Galactic WR stars, the NEOWISE-R database was examined at 
the positions of the WC stars in the Large Magellanic Cloud (LMC). Confusion 
was a significant limitation here, but useful data could be retrieved for 
12 WC stars in less crowded regions. 
The LMC stars are listed in Table~\ref{TLMC}, by BAT99 \citep*{BAT99} number, 
and giving the same photometric quantities as in Table~\ref{TSum}.

% beam sizes from Wright et al. (2010), section 3

\begin{figure}                                        % -------- Histograms
\centering
\includegraphics[width=8cm]{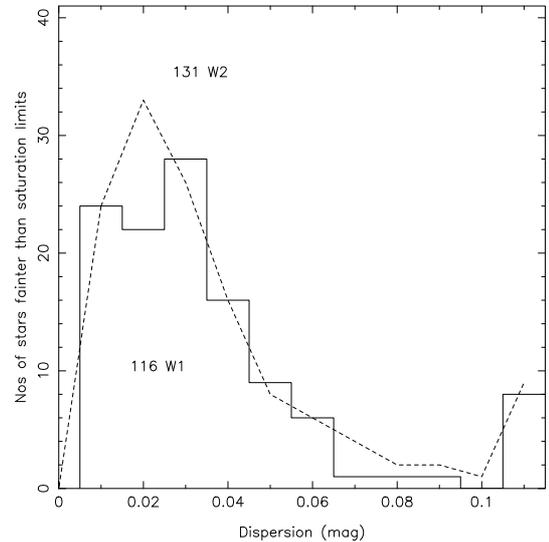}
\caption{Frequency distributions of dispersions in unsaturated $W1$ (histogram) 
and $W2$ (broken line). The final bins are $\sigma > 0.1$.}
\label{Fhist}
\end{figure}

Individual $W1$ and $W2$ `A'-quality profile-fitted magnitudes were retrieved 
from the Single Exposure Source Tables, excluding observations from bad frames 
or found more than 1~arc sec from the nominal position. 
No hard limit was set for the profile fitting metric, $\chi^2$, but individual 
observations having significantly greater ($\sim 5\times$) values than than 
those of other observations in the same data set were also excluded. 
This allowed searching for variability by stars having a neighbour always 
affecting the profile fitting in the same way and not introducing variability 
to the derived magnitudes.

The individual $W1$ and $W2$ were examined for short-term systematic variation 
within each visit. This was found in one case only: WR\,60-3 showed apparent 
eclipses of $\Delta W1 \simeq 0.25$ mag. lasting less than a day in several of 
the 1.25-d  NEOWISE-R visits, and is discussed in Section \ref{S60_3}. 
In order to search for long-term variations, the individual magnitudes in each  
visit which included at least five acceptable observations were averaged to 
provide $W1$ and $W2$ corresponding the averaged date of that visit, thereby 
building a set of (usually) ten\footnote{There was a hiatus in the observing in 
early 2014, so that stars between WR\,121-4 and WR\,125 in Table~\ref{TSum} were 
observed in nine visits only and the timing of the mission allowed 11 visits for 
a few stars.} independent observations spaced by approximately six months for each 
star. These averages form the datset for the present investigation. 
Owing to the high ecliptic latitude ($\beta \sim -86\degr$) of the LMC and 
greater overlap of the survey strips, the stars received 5--10 times as many 
observations per visit as the Galactic stars, so that the accuracies of the 
average $W1$ and $W2$ are comparable to those of the Galactic stars.
Their means and dispersions are collected in Tables~\ref{TSum} and \ref{TLMC}; 
they do not include measures in the All-Sky, 3-Band Cryo or NEOWISE Post-Cryo 
surveys of 2010, which are kept separate, but were 
subsequently compared with the NEOWISE-R data in the examination of variability.
Given the focus of the present study on long-term variability, and in order 
to reduce the influence of intra-visit variability, whether intrinsic like that 
of WR\,60-3 referred to above, or observational, an additional variability 
metric was introduced. This is the ratio, $R$, of the inter-visit (semestrial) 
dispersion to the mean intra-visit (orbital) dispersion. These are also given in 
Tables \ref{TSum} and \ref{TLMC}.

It is apparent from Tables \ref{TSum} and \ref{TLMC} that the dispersions in 
$W1$ and $W2$ are generally small (Fig.\,\ref{Fhist}); the median dispersions  
are 0.026 and 0.022 mag. (respectively) for the sources fainter than the 
saturation limits. 
From inspection of the distribution of dispersions, candidate variables 
are taken to be stars having dispersions $\sigma > 0.05$ and semestral to 
orbital dispersion ratios $R > 1.5$. 
The stars about 1--2~mag.\ brighter than the saturation limits, having 
$W1 = 6-8$ and $W2 = 6-7$, show slightly greater dispersions, 0.037 and 
0.031 mag.\ in $W1$ and $W2$ but can still be used to search for variability, 
albeit with a higher threshold: $\sigma > 0.08$. 

Evidently, the variability of WC stellar wind emission is small and the 
question arises: which of the few stars found to be variable owe this to 
varying dust emission?

Three stars previously known to be variable can be discounted. Two, WR\,9 and 
WR\,30, are optically observed atmospheric eclipsing binaries \citep{Lamontagne96} 
and the third, WR\,143, was observed to vary in $JHK$ by \citet{Watson143}, who 
ascribed the variation to the Be companion they identified. 
They are not dust makers and will not be considered further here.

The stars found to be variable are examined individually in 
Section~\ref{Sindividual} for periodic or systematic variation 
suggestive of a CWB, but first we consider the question of whether the stars, 
including the apparently non-variable sources, are dust makers.

%------------------------------------------------------------------------------
\section{Search for dust emission} % ------------------------------------------
\subsection{Interstellar reddening}
%------------------------------------------------------------------------------

Distinguishing the SEDs of heated dust from the free-free emission of the 
stellar winds requires not only observations at IR wavelengths sensitive 
to dust emission, but also knowledge of the interstellar reddening. This is  
important for distant WR stars in the Galactic Plane because the effect of 
heavy reddening on the near-IR colours can mimic that of Planckian dust emission. 
For WR stars whose reddening can be determined from optical photometry, 
uncertainties in the optical reddening translate into much ($\sim 10 \times$) 
smaller uncertainties in the IR, so that the effects of dust emission and 
reddening on the IR SED are almost orthogonal. Of the stars in 
Tables~\ref{TSum} and \ref{TLMC}, about one quarter have $b$ and $v$ 
on the narrow-band system optimised for WR stars \citep{SmithNB}, from 
aperture photometry or calibrated spectra (e.g. \citet{SMSNPL}). Most of 
the recently discovered WR stars are too heavily reddened, however, to have 
such data so it is necessary to look to photometry at slightly longer 
wavelengths, but still not affected by dust emission. 

Most of the brighter southern hemisphere WR stars in the sample have $i$ 
band photometry in the DENIS (Deep Near Infrared Survey of the Southern Sky, 
\citet{DENIS}) survey. This is currently being superseded by $i$ magnitudes 
from the ongoing VST Photometric Survey, VPHAS+, \citet{VPHAS}, and these 
$i$ (often together with $r$) data from DR2 and DR3 were used where possible. 
In the northern hemisphere, $i$ and $r$ photometry measured 
in the INT/WFC Photometric H$\alpha$ Survey (IPHAS, \citet{IPHAS}) 
were sought, taking data from the IPHAS2 \citep{IPHAS2}. Owing to their 
slightly different wavelengths (0.79~$\micron$ vs. 0.77~$\micron$), the 
DENIS and IPHAS/VPHAS $i$ magnitudes were treated separately for 
de-reddening and conversion to monochromatic fluxes.

Some of the southern stars ($\delta < -20\degr$) have been observed in the 
$Z$ (0.88 $\micron$) and $Y$ (1.02 $\micron$) bands in the VISTA Variables 
in the V\'{i}a L\'actea (VVV) survey \citep{VVV}. Data were retrieved 
from Data Release 4 in the VISTA Science Archive \citep{VSA}. 

% {\tt Note: searched DECaPS \citet{DECaPS} (Southern MW) for shorter 
% wavelength photometry of WR60-4, WR75-11 but not found; 
% WR91-1, z and Y (3,3 OK) found. WR60-1 and WR75-15 have DECam Z 
% but DECam Z filter is much wider 0.85-1.00 micron than VVV so don't use!}

%------------------------------------------------------------------------------
\subsection{The SEDs}
\label{SSEDs}
%------------------------------------------------------------------------------

To form the SEDs, $JHK_s$ data were taken from the Two Micron All Sky Survey 
(2MASS, \citet{2MASS}) or, for the more heavily reddened stars whose tabulated 
2MASS magnitudes are upper limits ({\tt ph\_qual = U}), the UKIRT Infrared Deep 
Sky Survey (UKIDSS, \citet{UKIDSS}) Galactic Plane Survey (GPS, \citet{GPS}) 
or the VVV. In the mid-IR, the {\em WISE} phtotometry was augmented with 
[3.6]. [4.5], [5.8] and [8.0] magnitudes observed in the {\em Spitzer}  
Galactic Legacy Infrared Mid-Plane Survey Extraordinaire 
(GLIMPSE, \citet{GLIMPSE1,GLIMPSE2}). 

All data were de-reddened using the `Wd1+RCs' reddening law determined by 
\citet{Damineli}, duly adjusted for the wavelengths of photometric bands 
used in the present study.
As pointed out by \citet{Damineli} and references therein, there are real 
differences in the reddening laws in different directions in the Galaxy, on 
small and large scales, and it is probable that many of the stars in this 
study are sufficiently heavily reddened for this to be an issue; but not 
one that can be addressed with the data presently available. 
The dust-free continua were inially assumed to follow a power law
$\lambda F_{\lambda} \propto \lambda^{-1.96}$ following \citet{Morris}, 
who found the UV--1-micron continua of single WR stars to be well fit by        % $\sim 0.1 - 1.0-\mu$m 
power laws, which were found to extend into the IR by \citet{Mathis}.
For each star, $A_V$ was then determined by fitting the shortest wavelength 
data available; if its continuum followed a different spectral index from that 
adopted (Morris et al. found a dispersion $\sigma = 0.14$ in spectral indices 
for WC stars), this would lead to an incorrect $A_V$ but the effect in the 
IR would be smaller.
If the WC star has an OB companion, as in a CWB, the SED will be steeper 
at shorter wavelengths depending on the relative contributions of the two 
components, and the effects of this will be discussed below.

Determination of the SEDs of the apparently non-varying WC stars revealed a 
number of previously unidentifed dust emitters, which are listed in 
Table~\ref{Tnewdust}. Their spectral types, from the Galactic WR Catalogue, 
are mostly WC8--9, as expected, but three stars, WR\,102-22, WR\,111-10 (WC7) 
and WR\,124-10 (WC6) have earlier types, making them the first WR stars 
having spectral subtypes earlier than WC8 to be identified as apparently 
non-variable persistent dust makers.

\begin{table}                     % -------------------------- Table {Tnewdust}
\centering
\caption{Persistent, apparently constant, dust-makers, newly identified 
as such in this study.
The spectral types are from the Galactic WR Catalogue and should now have `d' 
appended. The column `dust/wind' gives the ratio of dust to wind emission 
at 3.4~$\micron$, $T_{dust}$ is that at the inner edge of the cloud 
(isothermal in the case of WR\,111-10), r.m.s. gives the quality of the 
fit to the data while `bands' indicates the photometric bands used for 
determinations of $A_V$ and fitting the stellar wind continuum.}
\label{Tnewdust}
\begin{tabular}{rlrrcrl}
\hline
WR     & Type & dust/ & $T_{dust}$  & r.m.s.& $A_V$ & bands \\
       &      & wind  &             & mag.  &       &       \\ 
\hline
 60-2  & WC8  &  5.4 & 1177$\pm$28  & 0.11  & 15.0  & $iZYJ$ \\ % 0.11; <Tg> 958, 0.15, no Gaia pi
 75-7  & WC9  & 30.0 & 1165$\pm$49  & 0.14  &  4.1  & $riZY$ \\ % .139; <Tg> 1055, 0.142, no Gaia pi 
 75-15 & WC8  &  1.7 & 1027$\pm$67  & 0.08  & 20.0  & $YJ$   \\ % .079; <Tg> 914, 0.083, not in Gaia
102-22 & WC7  & 11.5 & 1418$\pm$67  & 0.12  &  9.5  & $riZY$ \\ % fit 1.2-12, r/r_0  = 3
111-10 & WC7  & 44.8 &  818$\pm$34  & 0.11  &  5.4  & $riiJ$ \\ % isothermal best, 0.113, pi .935+/-.413
122-1  & WC8  &  2.2 & 1083$\pm$99  & 0.10  &  4.4  & $riiJ$ \\ % 0.099; <Tg> 914, 0.11, no Gaia pi
123-4  & WC8  & 21.6 &  953$\pm$53  & 0.15  &  9.5  & $izJ$ \\ % use PanSTARRS z, J, not in Gaia
124-10 & WC6  & 12.8 & 1491$\pm$72  & 0.14  &  8.1  & $ri$  \\ % 10x IPHAS DR2, no Gaia pi
124-20 & WC9  &  4.4 & 1012$\pm$68  & 0.13  & 11.7  & $iJ$  \\ % .129; <Tg> 859, 0.148, no Gaia pi 
124-22 & WC9  &  2.7 &  980$\pm$70  & 0.08  & 13.9  & $iJ$  \\ % no Gaia pi
\hline
\end{tabular}
\end{table}

\begin{figure}                                        % -------- 60-2 & 111-10 SEDs
\centering
\includegraphics[width=8cm]{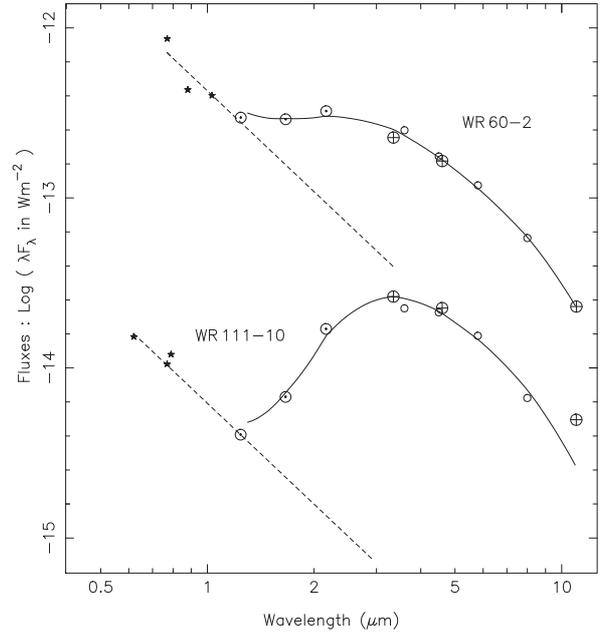}
\caption{SEDs of two of the newly identified persistent dust makers. Fluxes 
from $i$, $Z$ and $Y$ (WR\,60-2) $r$ and $i$ (both VPHAS+ and DENIS, the latter 
brighter but more uncertain) are marked $\star$; those from $J$, $H$ and 
$K_s$ marked $\odot$, those from {\em WISE} data marked $\oplus$ and those 
from GLIMPSE data marked $\circ$. Broken lines represent the wind continua 
fitted to the short wavelength photometry and solid lines the wind+dust models 
fitted to the IR data, excluding $W3$ in the case of WR\,111-10.}
\label{FperSEDs}
\end{figure}

The dust emission was modelled using clouds of amorphous carbon grains 
assumed to have optical properties of the `ACAR' laboratory grains studied 
by \citet{ColangeliAC}. The absorption coefficients were calculated from 
the optical properties for this sample given by \citet{ZubkoACAR}. 
The emission was assumed to be optically thin, with dust density falling 
off radially as $r^{-2}$, appropriate for dust formed in a stationary wind 
with constant mass-loss rate. 
The grain temperature is determined by radiative equilibrium in the stellar 
radiation field, falling off as $T_g \propto r^{-0.4}$ following \citet{WHT}. 
The temperature of the dust nearest the star and the amount of dust were 
found by fitting the observed fluxes assumed to define a representative 
SED for the non-variable stars although different wavelength regions were 
observed at different times.  
The radial extent of the cloud is poorly constrained by the data as the 
more distant dust is too cool to contribute significant emission at the 
wavelengths observed. In practice, the quality of fits to the data by 
successive models having increasing radial extent were compared and 
the extension terminated when there was no improvement to the fit. 
Conversion of IR flux to dust mass requires knowledge of the distance to 
the source, which enters as its square. Distances to these stars are poorly 
constrained: only one of the stars (WR\,111-10) has a parallax in Gaia DR2 
\citep{GaiaDR2} and that is very uncertain ($\sigma_{\varpi}/\varpi > 0.4$). 
Determination of distances from the de-reddened photometry requires 
knowedge of the stars' absolute magnitudes, which are very uncertain in the 
light of the recent study of the luminosities of WC stars having Gaia 
parallaxes \citep{SanderGaia} showing them to be very dispersed, as well 
as the possible presence of undetected luminous companions to the WC stars.
For these reasons, the amount of circumstellar dust is expressed as the ratio 
of dust to wind emission at a reference wavelength, that of the $W1$ filter. 
These ratios, the dust temperatures, values of $A_V$ and photometric bands 
used for their determination are given in Table~\ref{Tnewdust}. 

The de-reddened SEDs of the variable sources were then examined for evidence 
of dust formation, for which the criterion was taken to be `excess' flux of 
at least 10 per cent of the wind flux in the 2--4-$\mu$m region. 
It is possible that some of the less intense dust emitters have been missed 
but, in most cases, the dust emission was so much brighter than the wind that 
the classification as dust  makers is unequivocal.

\begin{table}                 % ------------------------------- Table {Tnewvar}
\centering
\caption{Galactic WC stars newly identified as variable dust-makers. The 
amplitudes $\Delta W1$ are based on all the {\em WISE} data except for WR\,77t, 
which is bright enough for saturation to affect the NEOWISE-R data, so its 
$\Delta W1$ comes from the All-Sky and 3 Band Cryo observations. 
Also given are the extinctions, $A_V$. The columns `dust/wind' and `Epoch' 
give the ratio of dust to wind emission at 3.4~$\micron$ and epoch at which 
it was measured while the final column gives the type of variability (see text).}
\label{Tnewvar}
\begin{tabular}{rlrrrll}
\hline
WR     & Type & $\Delta W1$ & $A_V$ & dust/ &  Epoch  & Var. \\
       &      &             &       &  wind &         &      \\
\hline
 46-7  & WC6-7 & 0.86   & 10.5      &  5.3  & 2010.57 & V    \\ %1040+/-50, no Gaia pi
 47c   & WC5   & 0.41   &  5.4      &  0.4  & 2010.10 & Ep   \\ % 980+/-40, pi .029+/-.030
 60-4  & WC8   & 0.32   & 26        &  0.7  & 2014.12 & Ep   \\ %1010+/-160, not in Gaia
 75aa  & WC9d  & 0.17   &  4.7      &  9.5  & 2010.17 & V    \\ % 920+/-70, no Gaia pi
 75d   & WC9   & 0.32   &  5.5      &  1.3  & 2010.67 & V    \\ % 920+/-110, no Gaia pi
 75-11 & WC9d? & 0.24   & 18.1      &  0.2  & 2010.67 & Ep    \\ % 885+/-260
 77t   & WC9d  & 0.49   &  8.9      &  1.8  & 2010.67 & V    \\ % 860+/-40, not in Gaia
 91-1  & WC7   & 0.66   & 27        &  0.2  & 2014.20 & Ep    \\
122-14 & WC8   & 0.25   & $\lesssim$ 48 & $\gtrsim$ 0.3  & 2010.75 & Ep?\\
125-1  & WC8   & 0.33   &  5.1      &  0.2  & 2010.30 & Ep    \\ % 560+/-100
\hline
\end{tabular}
\end{table}

Galactic variable dust makers are listed in Table~\ref{Tnewvar}, which includes  
stars previously identified as dust makers or uncertain (`d' or `d?') and 
those newly found to be dust makers here. Along with amplitudes $\Delta W1$, 
the Table gives the reddening determined for each and ratio of dust to wind 
flux in the $W1$ band, susceptible to uncertainty due to variability. 
For one of the variables, WR\,122-14, it is not possible with data presently 
available to distinguish between the effects of very heavy reddening and 
dust emission: deeper photometry at one or more wavelengths shorter than 
$J$ is required to disentangle these effects (cf. Section~\ref{S122-14}). 
The final column aims to classify the variation as episodic (Ep), in which 
dust is formed for only part of the time, or variable (V), in which dust forms 
persistently but at a variable rate. Stars which show stellar wind emission 
some of the time can securely be identified as episodic dust makers but the 
distinction is otherwise more difficult, as is discussed further below. 
To the stars in this Table must be added the episodic dust maker HD~38030 
in the LMC. Details are given in Section~\ref{Sindividual}, where the stars 
are discussed individually.  

Four of the stars newly found to be dust makers (Table~\ref{Tnewdust}), WR\,60-2, 
122-1, 124-20 and 124-22, and not considered variable on the basis of their 
dispersions, have semestral-to-orbital dispersion ratios $R > 1$ (Table \ref{TSum}) 
and may be low amplitude variables better included in Table \ref{Tnewvar}, but this 
needs further study.

Of the stars classified as possible dust makers (`d?') in the Catalogue, 
one, WR\,75-11, was found to be a variable dust maker and is included in 
Table \ref{Tnewvar}. The SEDs of the two other stars in the sample classified 
as possible dust makers, WR\,72-3 and WR\,124-5, were examined and found to be 
fitable by stellar winds suffering 
extinctions of $A_V \simeq$ 23.4 and 22 mag., respectively, with no evidence 
for dust emission. 
It is striking that all of the stars found to be variable in the NEOWISE-R 
data, with the exceptions of WR\,60-3 and the previously known variables 
mentioned above, are dust makers.

%------------------------------------------------------------------------------
\section{Comments on individual dust-emission variables}
\label{Sindividual}
%------------------------------------------------------------------------------

\subsection{The known episodic dust-maker WR\,19}              % ------- WR 19

\begin{table}
\centering
\caption{Photometric history of WR\,19 from {\em WISE} observations. The phases 
are on the elements of \citet{W19orb}.}
\label{T19}
\begin{tabular}{llccl}
\hline
Date     & Phase & $W1$ & $W2$ & Survey \\
\hline
2010.01  & 0.28  & 8.15$\pm$0.01 & 7.54$\pm$0.01 &  All-Sky \\
2010.50  & 0.32  & 8.20$\pm$0.01 & 7.70$\pm$0.01 &  All Sky \\
2011.02  & 0.37  & 8.22$\pm$0.01 & 7.78$\pm$0.01 &  Post Cryo \\
2014.02  & 0.67  & 8.25$\pm$0.01 & 7.93$\pm$0.01 &  NEOWISE-R \\
2014.51  & 0.72  & 8.25$\pm$0.01 & 7.91$\pm$0.01 &  NEOWISE-R \\
2015.02  & 0.77  & 8.25$\pm$0.01 & 7.92$\pm$0.01 &  NEOWISE-R \\
2015.49  & 0.82  & 8.25$\pm$0.01 & 7.93$\pm$0.01 &  NEOWISE-R \\
2016.02  & 0.87  & 8.26$\pm$0.01 & 7.92$\pm$0.01 &  NEOWISE-R \\
2016.48  & 0.91  & 8.24$\pm$0.01 & 7.92$\pm$0.01 &  NEOWISE-R \\
2017.01  & 0.97  & 8.24$\pm$0.01 & 7.92$\pm$0.01 &  NEOWISE-R \\
2017.46  & 0.01  & 8.15$\pm$0.01 & 7.74$\pm$0.01 &  NEOWISE-R \\
2018.01  & 0.07  & 5.50$\pm$0.08 & 4.51$\pm$0.09 &  NEOWISE-R \\ %  sigma_m 0.08, 0.09
2018.45  & 0.11  & 6.02$\pm$0.05 & 5.07$\pm$0.12 &  NEOWISE-R \\ %  sigma_m 0.05  0.12
\hline
\end{tabular}
\end{table}

\begin{figure}
\centering
\includegraphics[angle=270,width=8.5cm]{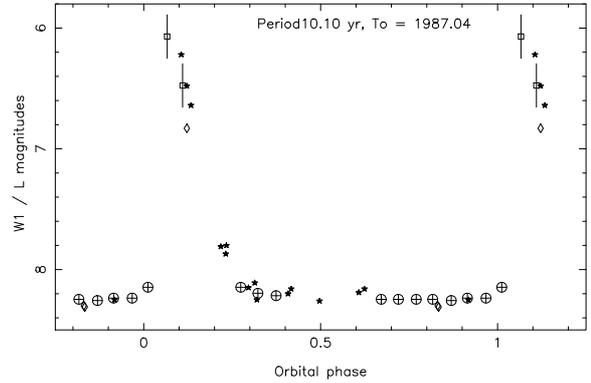}
\caption{Light curves in $W1$, $\oplus$ marking the 2010-17 observations and 
$\Box$ the saturation-adjusted 2018 data (see text, the error bars include the 
uncertainties in the offsets) compared with those in $L^{\prime}$ ($\star$) and 
$L$ ($\Diamond$) from the earlier studies referred to.}
\label{F19}
\end{figure}

The first IR photometry of WR\,19 in 1988-90 showed it to be fading from an 
inferred dust-formation episode \citep{W19}. Further observations by 
\citet{Veen19} found another dust formation epsiode, from which they derived 
a period of 10.1~y. They also observed absorption lines in its spectrum, 
indicating the presence of a O9.5-9.7 type companion to the WC star, in line 
with the original classification WC5+OB by \citet{Smith68} in her discovery paper. 
Fading from a  third dust-formation episode was observed by \citet*{W19orb}, 
who also determined a RV orbit (using the IR period as a prior) from the 
absorption lines showing high eccentricity ($e = 0.8$) and having periastron 
passage close to the time of dust formation.

The earliest {\em WISE} observations of WR\,19 made in 2010 in the All-Sky 
Survey (in two visits, separated by six months) show WR\,19 to be still fading 
from its 2007--08 dust outburst. The All-Sky Survey data in Table~\ref{T19} 
come from averaging the individual observations in the Single 
Exposure Source Table for the two visits separately in order to retain the 
temporal information. When observed in the Post-Cryo survey in 2011, the IR 
emission had faded to close to the average wind level. 

The first seven NEOWISE-R observations (Table~\ref{T19}), 
which cover orbital phases 0.67--0.97 on the elements of \citet{W19orb}, 
show constant values of $W1$ = 8.25 and $W2$ = 7.92, 
These are consistent with the stellar wind flux $L^{\prime} = 8.20$ and 
$L = 8.30$ measured at the nearby wavelengths of 3.8 and 3.6~$\micron$ 
and $M \simeq 7.7$ measured at 4.7~$\micron$ \citep{W19,Veen19}, and give 
a better definition of the stellar wind flux level and its constancy than 
the ground-based data. 
The eighth NEOWISE-R observation at phase 0.01 shows $W1$ and $W2$ to have 
brightened by 0.09 and 0.18 mag. respectively, signalling the beginning of 
a new dust formation episode. 
The 2018 observations, at phases 0.07 and 0.11, show $W1$ and $W2$ to have 
brightened to above the NEOWISE-R saturation limits. To get estimates of 
the de-saturated $W1$ and $W2$ for comparison with the earlier data, 
offsets from \citet[figure 6]{NEOWISER} were applied to these data.
The $W1$ photometry, together with the earlier $L$ and $L^{\prime}$ 
\citep{W19,Veen19}, is plotted against orbital phase in Fig. \ref{F19}.
The NEOWISE-R data fit the earlier data well and give a better idea of 
the timing of the dust formation. Finer cadence than that provided by the 
NEOWISE-R observations will be needed to define the rise to maximum more 
precisely, but it is evident that the duration of dust formation is very 
brief -- presumably 
related to the relatively high eccentricity ($e = 0.8$, \citet{W19orb}), 
making WR\,19 an analogue of WR\,140. It deserves a better, double-lined 
spectroscopic orbit and search for spectroscopic signatures (`sub-peaks' 
on low excitation lines) of colliding wind effects.

\citet*{Sugawara19125} have observed the X-ray emission from WR\,19, 
finding that, as it approached periastron, the column density increased, 
as expected if the colliding wind X-ray source moved more deeply into 
the WR wind. 
\citet*{LeithererRadio} observed only upper limits to the radio flux from 
WR\,19 but deeper observations to look for non-thermal radio emission would 
be worthwhile.

%---------------------------------------------------------------------
\subsection{WR\,46-7 = 2MASS J12100795--6244194}  % ------------------
%---------------------------------------------------------------------

\begin{table} 
\centering 
\caption{IR Photometric history of WR\,46-7.  The magnitudes tabulated 
under $W1$ and $W2$ for the GLIMPSE observations are [3.6] and [4.5]  
and the date comes from the Spitzer Heritage Archive.} 
\label{T46_7}
\begin{tabular}{lrrrrl}
\hline
  Date  &$K_s$ &  $W1$ & $W2$ & $W3$ & Source\\
\hline

2000.08 &  9.84 &       &      &      & DENIS  \\
2000.27 &  9.74 &       &      &      & 2MASS \\  % April 8
2004.55 &       & 8.29  & 7.85 &      & GLIMPSE\\ 
2010.08 &       & 9.19  & 8.44 & 6.40 & All-Sky \\
2010.57 &       & 8.32  & 7.67 & 6.28 & All-Sky \\
2011.08 &       & 9.06  & 8.41 &      & Post-Cryo \\
2013.11 & 10.40 &       &      &      & VVV \\
2013.11 & 10.35 &       &      &      & VVV \\
2013.23 & 10.36 &       &      &      & VVV \\
2014.09 &       & 9.09  & 8.43 &      & NEOWISE-R \\
2014.57 &       & 9.18  & 8.49 &      & NEOWISE-R \\
2015.08 &       & 8.33  & 7.68 &      & NEOWISE-R \\
2015.56 &       & 8.99  & 8.35 &      & NEOWISE-R \\
2016.07 &       & 9.12  & 8.41 &      & NEOWISE-R \\
2016.55 &       & 8.36  & 7.69 &      & NEOWISE-R \\
2017.08 &       & 8.99  & 8.34 &      & NEOWISE-R \\
2017.53 &       & 9.14  & 8.41 &      & NEOWISE-R \\
2018.08 &       & 8.39  & 7.68 &      & NEOWISE-R \\
2018.52 &       & 8.77  & 8.12 &      & NEOWISE-R \\
\hline
\end{tabular}
\end{table}

\citet{MV09} identified WR\,46-7 as a Wolf-Rayet star from its IR colours 
and classified it as a WC5--7 star from its $K$-band spectrum.
The NEOWISE-R observations (Table~\ref{T46_7}) show it to be $\sim 0.8$ mag. 
brighter in $W1$ and $W2$ in the third (2015.08), sixth (2016.65) and 
ninth (2018.08) visits. Magnitudes from the {\em WISE} All-Sky survey 
which, in this case, observed WR\,46-7 in two visits separated by six months, 
shows that it was similarly brighter in all three filters in the second 
(2010.57) All-Sky visit. 
The entries for $W1$, $W2$ and $W3$ in Table~\ref{T46_7} for the 
All-Sky Survey come from averaging the individual observations in the 
Single Exposure Source Table to separate those made in the two visits, as 
for the NEOWISE-R data, in order to retain the temporal information.
Six months later, observation in the Post-Cryo survey (2011.08) 
found WR\,46-7 in its faint state again.

When observed in the GLIMPSE survey, WR\,46-7 was also in its bright state. 
The [3.6] and [4.5] magnitudes from the GLIMPSE surveys are 
listed under $W1$ and $W2$ in Table~\ref{T46_7} and the Tables in the remainder 
of this Section without adjustment for the differences in the photometric bands. 
\citet{SpitzerWISE} found that $W1$ and $W2$ and the corresponding IRAC [3.6] 
and [4.5] magnitudes for a large sample of stars near the ecliptic poles 
agreed well, with small offsets resulting from slightly different wavelengths 
of the filters. 
In case the different continua and emission lines in WC spectra might give 
different results, the {\em WISE} and GLIMPSE magnitudes of WC stars were 
compared and found to show mean offsets $W1$--[3.6] = -0.02$\pm$0.03 and 
$W2$--[4.5] = 0.06$\pm$0.02 from 25 and 26 stars respectively. 
As these are smaller than the dispersions in the differences 
($\sigma$ = 0.14 and 0.12 mag.), the GLIMPSE data have not been adjusted for 
the light curves in this paper, but the {\em WISE} and GLIMPSE data sets are 
treated separately for the production of the SEDs. 

\begin{figure}                                  % -------- light curves WR 46-7
\centering
\includegraphics[angle=270,width=8.5cm]{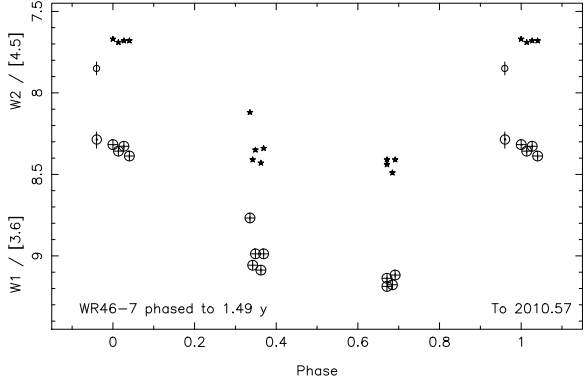}
\caption{Mid-IR light curves of WR\,46-7 using {\em WISE} $W1$ ($\oplus$) and
$W2$ ($\star$) and GLIMPSE [3.6] ($\odot$) and [4.5] ($\circ$) phased against 
a period of 1.49 y with zero phase set to the date of the second All-Sky 
observation in 2010.57. Error bars are $\pm 1 \sigma$.}
\label{Flc46_7}
\end{figure}

The observation date is taken from the {\em Spitzer} Heritage Archive. 
The interval between this date and that of the brighter All-Sky observation 
($\simeq 6$~y.) is four times the c. 1.5-yr period suggested by the {\em WISE} 
data. The GLIMPSE data are consistent with the latter, but do not improve the 
phase coverage. The data are too few and evenly spaced for a confident period 
determination: Lafler-Kinman \citep{LaflerKinman} searches on the {\em WISE} 
and GLIMPSE data give periods of 1.49~y, unfortunately close to three times 
the semesterly cadence of the {\em WISE} visits, and 0.75~y. Data having a 
different cadence are required to determine the period; unfortunately, 
WR\,46-7 is too bright to get a $K_s$ light curve from the VVV (most of 
the $K_s$ observations are flagged as being near saturation) so, for the 
present, the 1.49-y period is adopted. The mid-IR light curves phased to 
this period are plotted in Fig.\,\ref{Flc46_7}.

\begin{figure}          % ----------------------------------------- SED WR 46-7
\centering
\includegraphics[width=8.5cm]{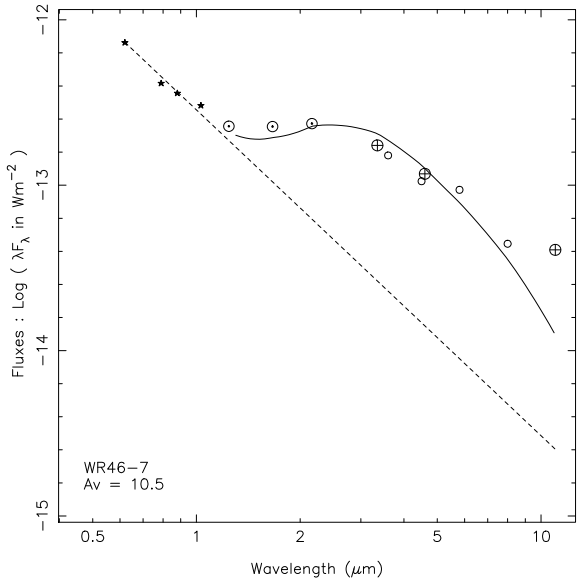}
\caption{SED of WR\,46-7 at its brightest, using $W1$ and $W2$ observed 
in 2010.08 (marked $\oplus$) and the GLIMPSE [3.6], [4.5], [5.8] and 
[8.0] observed in 2004.55, close to the same phase, marked $\circ$), 
and $JHK_s$ from 2MASS (marked $\odot$). Also shown, but not used for 
the dust model, is $W3$.
The short wavelength data used to fit the wind continuum are 
VPHAS+ DR2 $r$ and $i$ and VVV $Z$ and $Y$ (all marked $\star$).}
\label{F46-7SED}
\end{figure}

The SED representing WR\,46-7 near maximum is given in Fig.\,\ref{F46-7SED}. 
The line is the flux from a model comprising wind emission fitted to short 
wavelength data ($r$, $i$, $Z$ and $Y$) together with a cloud of 1040-K dust.
The longer wavelength GLIMPSE data, [5.8] = 7.52 and [8.0] = 7.36, observed 
at phase 0.92 shortly before maximum are consistent with strong dust emission. 
The fit excludes $W3$ (11~$\mu$m), which has a significantly higher 
profile-fitting metric $\chi^2$, and which may owe its anomalous brightness 
to the inclusion of extended emission not associated with the WR star.
The model for the wind emission was a power-law as described in 
Section~\ref{SSEDs} above; because WR\,46-7 is a good candidate CWB, the fit was 
repeated using an alternative SED derived from the WC8+O9 binary $\gamma$~Velorum
yielding a slightly higher reddening ($A_V$ = 10.8 compared with 10.5) and 
3.4-$\mu$m dust/wind ratio (8.5 compared with 5.3, cf. Table~\ref{Tnewvar}). 
This serves to illustrate the effect of uncertainty in the wind emission; 
without knowledge of the relative contributions of the possible OB companion 
and WC6-7 star fluxes, the power-law wind SEDs will be used for the present.
 
The mean $W1$--$W2$ colours near maximum, 0.67$\pm$0.02, and minimum (phase 0.70), 
0.73$\pm$0.02, show evidence for cooling of the dust as it is dispersed by the 
stellar wind. They are much redder than the reddened stellar wind colour, 0.37, 
suggesting that dust formation continues the whole time, albeit at a variable rate.

%------------------------------------------------------------------------------
\subsection{WR\,47c = SMSNPL 7}                            % ---------- WR 47c
%------------------------------------------------------------------------------

\begin{table} 
\centering 
\caption{mid-IR Photometric history of WR\,47c. }
 
\label{T47c}
\begin{tabular}{lccl}
\hline
  Date  &     $W1$      &    $W2$       & Source\\
\hline
2004.55 & 8.94$\pm$0.04 & 8.64$\pm$0.03 & GLIMPSE \\
2010.10 & 9.01$\pm$0.01 & 8.60$\pm$0.04 & All Sky \\
2010.59 & 9.00$\pm$0.03 & 8.61$\pm$0.04 & All Sky \\
2012.27 & 9.11$\pm$0.05 & 8.74$\pm$0.03 & Deep GLIMPSE\\
2014.10 & 9.29$\pm$0.01 & 8.95$\pm$0.01 & NEOWISE-R \\
2014.60 & 9.34$\pm$0.01 & 8.99$\pm$0.01 & NEOWISE-R \\
2015.10 & 9.38$\pm$0.01 & 9.05$\pm$0.01 & NEOWISE-R \\
2015.58 & 9.41$\pm$0.02 & 9.09$\pm$0.01 & NEOWISE-R \\
2016.10 & 9.52$\pm$0.01 & 9.19$\pm$0.01 & NEOWISE-R \\
2016.56 & 9.47$\pm$0.01 & 9.16$\pm$0.01 & NEOWISE-R \\
2017.10 & 9.61$\pm$0.01 & 9.30$\pm$0.01 & NEOWISE-R \\
2017.55 & 9.55$\pm$0.02 & 9.26$\pm$0.01 & NEOWISE-R \\
2018.10 & 9.63$\pm$0.01 & 9.33$\pm$0.01 & NEOWISE-R \\
2018.54 & 9.59$\pm$0.01 & 9.34$\pm$0.02 & NEOWISE-R \\
\hline
\end{tabular}
\end{table}

\begin{figure}               % ---------------------------- light curve WR 47c
\centering
\includegraphics[width=8.5cm]{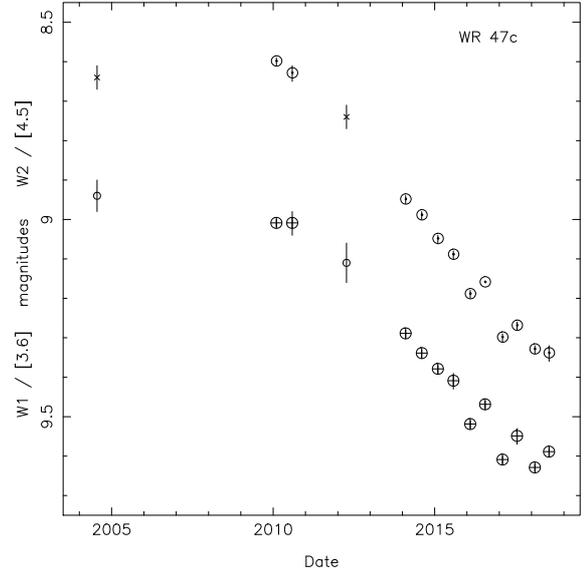}
\caption{Synoptic photometry of WR\,47c in $W1$ ($\oplus$), $W2$ ($\odot$), 
GLIMPSE [3.6] ($\circ$) and [4.5] ($\times$). Error bars are $\pm 1\sigma$.}
\label{Fsynoptic47c}
\end{figure}

WR\,47c was identified as a WR star in the narrow-band optical (4686-\AA) 
survey of \citet{SMSNPL}, who classified it as WC5 and gave magnitudes 
$b$ = 17.49, $v$ = 16.09 (on the narrow-band system \citep{SmithNB}) measured 
from their fluxed spectra. 
The NEOWISE-R observations (Table~\ref{T47c} and Fig.\,\ref{Fsynoptic47c}) 
show slow fading in 2014--2018 and are significantly fainter than in the 
two visits of the All-Sky Survey made in 2010.10 and 2010.59. The 2018 
$W1$ and $W2$ are constsnt with a reddened stellar wind fitted to the 
$b$, $v$, $r$, $i$, $Z$ and $Y$ photometry, classifying WR\,47c as 
an episodic dust maker.

The Deep GLIMPSE observations, made between the All-Sky and NEOWISE-R 
observations, fits the steady fading well (Fig.\,\ref{Fsynoptic47c}).
The 2004 GLIMPSE data, taken 5.5 years before the first All-Sky observation, 
also has WR\,47c bright.
The long duration of the fading observed with {\em WISE} suggests that there 
would not have been time for another dust formation and fading episode in the 
five years between the GLIMPSE and All-Sky observations. 
This implies that the system was at broad maximum in 2004--2010 and has a 
long period, exceeding 14~y.

%------------------------------------------------------------------------------
\subsection{Short-term variability of WR\,60-3 = MDM 11}      % -----   WR 60-3
%-----------------------------------------------------------------------------
\label{S60_3}

\begin{figure}
\centering
\includegraphics[angle=270,width=8.5cm]{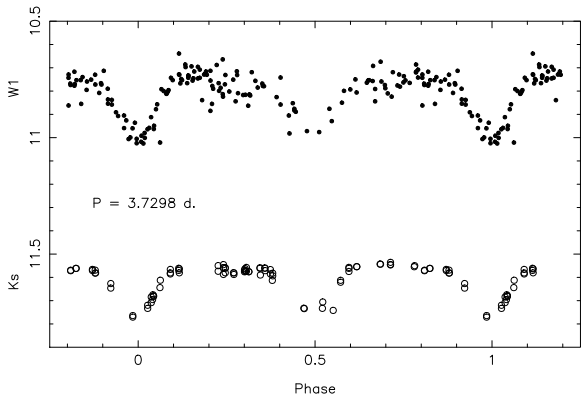}
\caption{Light curves in $W1$ and $K_s$ of WR\,60-3.}
\label{F60-3}
\end{figure}

As noted above, eclipses were found in the NEOWISE-R photometry of WR\,60-3. 
Examination of the 3-Band Cryo Single Exposure Source table showed another 
eclipse. A Lafler-Kinman period search on all the $W1$ photometry gave a 
well-defined period of 1.8649~d. 
Photometry in the VVV was searched for periodicity, the 80 $K_s$ magnitudes 
gave $P$ = 1.8648 days. 
Light curves phased to 1.8649~d show no convincing secondary minima but 
curves (Fig.\,\ref{F60-3}) phased to twice this period appear to show slightly 
different primary and secondary minima, so it is likely that the true period 
is 3.7298~d. There is no evidence for dust emission so this 
system will not be discussed further here.

%------------------------------------------------------------------------------
\subsection{WR\,60-4 = MDM 12}                                 % ------ WR 60-4
%------------------------------------------------------------------------------

\begin{table} 
\centering 
\caption{mid-IR Photometric history of WR\,60-4.} 
\label{T60_4}
\begin{tabular}{lccl}
\hline
  Date  & $W1$  & $W2$   & Source\\
\hline
2004.19 &  9.50$\pm$0.03 &  8.89$\pm$0.05  & GLIMPSE \\ % e[3.6] 0.03, e[4.5] 0.05
2010.12 &  9.74$\pm$0.02 &  8.91$\pm$0.02  & All-Sky \\
2010.61 &  9.74$\pm$0.02 &  8.89$\pm$0.02  & 3-Band Cryo \\
2014.12 &  9.52$\pm$0.02 &  8.65$\pm$0.02  & NEOWISE-R \\
2014.62 &  9.79$\pm$0.02 &  8.96$\pm$0.04  & NEOWISE-R \\
2015.12 &  9.69$\pm$0.02 &  8.88$\pm$0.02  & NEOWISE-R \\
2015.60 &  9.47$\pm$0.01 &  8.58$\pm$0.02  & NEOWISE-R \\
2016.12 &  9.68$\pm$0.02 &  8.86$\pm$0.03  & NEOWISE-R \\
2016.59 &  9.66$\pm$0.02 &  8.82$\pm$0.01  & NEOWISE-R \\
2017.12 &  9.48$\pm$0.03 &  8.59$\pm$0.02  & NEOWISE-R \\
2017.57 &  9.69$\pm$0.02 &  8.87$\pm$0.01  & NEOWISE-R \\
2018.12 &  9.73$\pm$0.02 &  8.90$\pm$0.02  & NEOWISE-R \\
2018.57 &  9.64$\pm$0.02 &  8.77$\pm$0.02  & NEOWISE-R \\
\hline
\end{tabular}
\end{table}

\begin{figure}
\centering
\includegraphics[angle=270,width=8.5cm]{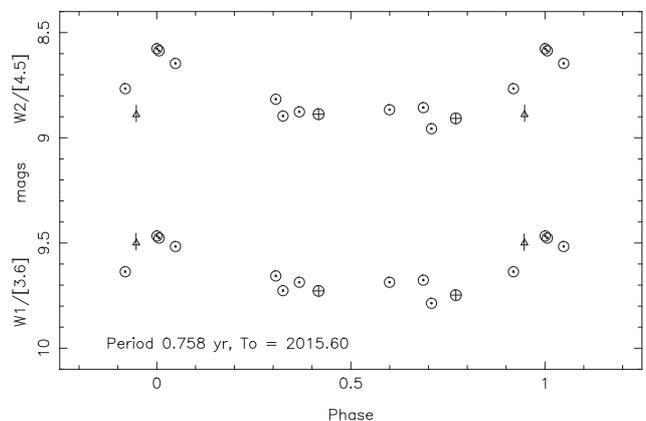}
\caption{Phased light curves in $W1$ and $W2$ ($\oplus$ All-Sky and 3-Band Cryo, 
($\odot$ NEOWISE-R) and GLIMPSE [3.6] and [4.5] ($\triangle$) of WR\,60-4.}
\label{F60-4}
\end{figure}

WR\,60-4 was discovered by \citet{MDM11}, who designated it MDM~12 and 
classified its spectrum as WC8. 
The photometry is collected in Table~\ref{T60_4}. Like WR\,46-7, the variations 
seem commensurate with the cadence of observations: the first, fourth and seventh 
NEOWISE-R observations show WR\,60-4 to be significantly brighter than in the 
All-Sky and 3-Band Cryo surveys, suggesting a period near 1.5~y, like WR\,46-7.
Period searches on these and the GLIMPSE photometry suggest periods near 1.55,  
0.76 and 0.38~d., examination of phased light curves (Fig.\,\ref{F60-4}) favours 
0.758~d. Further observations having a different cadence would help determine 
the period. 
There are 56 $K_s$ magnitudes observed in the VVV between 2010.4 and 2013.5 
which show small amplitude ($<$ 0.2 mag) variation but none were taken near 
the maxima derived from the mid-IR data nor do they favour any of the suggested 
periods.

The star is heavily reddened, with $A_V \simeq 26$ derived using $Y$ and $J$ 
from the VVV Survey. The stellar wind so reddened has $W1-W2 \simeq 0.85$. 
This is very close to the average $W1-W2$ = 0.83 observed between phases 
0.2 and 0.8, indicating no dust emission during this time and classifying 
WR\,60-4 as an episodic dust maker.  

% ------------------------------------------------------------------------------
\subsection{WR\,75aa = HBD 1}  % --------------------------------------- WR 75aa
%-------------------------------------------------------------------------------

\begin{table} 
\centering 
\caption{{\em WISE} Photometric history of WR\,75aa} 
\label{T75aa}
\begin{tabular}{lcccl}
\hline
  Date  &     $W1$       &     $W2$       &      $W3$      & Survey\\
\hline
2010.17 & 8.14$\pm$0.02  & 7.46$\pm$0.02  & 6.92$\pm$0.02  & All-Sky \\
2010.66 & 8.10$\pm$0.02  & 7.43$\pm$0.02  & 6.91$\pm$0.06  & 3-Band Cryo \\
2014.18 & 8.27$\pm$0.01  & 7.60$\pm$0.01  &                & NEOWISE-R \\ % fainter
2014.67 & 8.17$\pm$0.01  & 7.51$\pm$0.01  &                & NEOWISE-R \\
2015.17 & 8.18$\pm$0.01  & 7.51$\pm$0.01  &                & NEOWISE-R \\
2015.66 & 8.10$\pm$0.01  & 7.44$\pm$0.01  &                & NEOWISE-R \\
2016.17 & 8.15$\pm$0.01  & 7.49$\pm$0.01  &                & NEOWISE-R \\
2016.64 & 8.12$\pm$0.01  & 7.47$\pm$0.01  &                & NEOWISE-R \\
2017.17 & 8.17$\pm$0.01  & 7.50$\pm$0.01  &                & NEOWISE-R \\
2017.63 & 8.25$\pm$0.01  & 7.58$\pm$0.01  &                & NEOWISE-R \\ % fainter
2018.17 & 8.19$\pm$0.01  & 7.54$\pm$0.01  &                & NEOWISE-R \\
2018.61 & 8.10$\pm$0.01  & 7.44$\pm$0.01  &                & NEOWISE-R \\  
\hline
\end{tabular}
\end{table}

\begin{figure}
\centering
\includegraphics[angle=270,width=8.5cm]{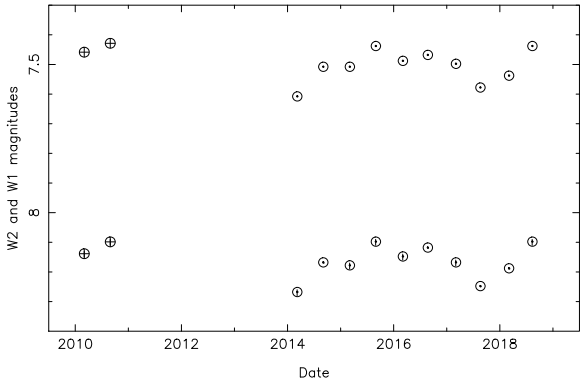}
\caption{Synoptic {\em WISE} All-Sky and 3-Band Cryo ($\oplus$) and NEOWISE-R 
($\odot$) $W1$ and $W2$ observed from WR\,75aa.}
\label{F75aa}
\end{figure}

\citet{HBD} identified WR\,75aa as a WC9 star from its strong line emission 
in the AAO/UKST H$\alpha$ survey, and as a dust emitter on the basis 
of its red and near-IR colours. The {\em WISE} photometry is collected in 
Table~\ref{T75aa} and plotted in Fig.\,\ref{F75aa}. 
Taken on their own, the NEOWISE-R data suggest a period near 3.25~y., but this 
is not fit by the 2010 data. The star is fainter than the NEOWISE-R saturation 
limits, so offsets between the NEOWISE-R and cryogenic (All-Sky and 3-Band Cryo) 
data are not expected and WR\,75aa has to be considered as an irregular variable. 
Unfortunately, there are no GLIMPSE data, presumably on account of the star's 
distance ($>2\degr$) from the Plane. Also, it is too bright to form a $K_s$ 
light curve from the VVV survey data.

% -----------------------------------------------------------------------------
\subsection{WR\,75d = HBD 3}   % --------------- Delta W1 = 0.32 p-p  -- WR 75d
% -----------------------------------------------------------------------------

\begin{table} 
\centering 
\caption{Mid-IR Photometric history of WR\,75d.} 
\label{T75d}
\begin{tabular}{lcccl}
\hline
  Date  &     $W1$       &       $W2$     & $W3$           & Source\\
\hline
2004.68 &  8.26$\pm$0.03 &  7.85$\pm$0.05 &                & GLIMPSE \\
2010.17 &  8.03$\pm$0.02 &  7.55$\pm$0.02 & 7.10$\pm$0.02  & All-Sky \\  % W4 = 6.91
2010.67 &  7.88$\pm$0.02 &  7.39$\pm$0.02 & 7.22$\pm$0.07  & 3-Band Cryo \\ %  eW3 0.07
2014.18 &  8.22$\pm$0.02 &  7.71$\pm$0.01 &                & NEOWISE-R \\
2014.68 &  8.24$\pm$0.02 &  7.73$\pm$0.02 &                & NEOWISE-R \\
2015.17 &  8.09$\pm$0.01 &  7.55$\pm$0.01 &                & NEOWISE-R \\
2015.66 &  8.13$\pm$0.03 &  7.56$\pm$0.03 &                & NEOWISE-R \\
2016.18 &  8.10$\pm$0.02 &  7.50$\pm$0.02 &                & NEOWISE-R \\
2016.64 &  8.23$\pm$0.02 &  7.69$\pm$0.01 &                & NEOWISE-R \\
2017.17 &  8.23$\pm$0.02 &  7.68$\pm$0.01 &                & NEOWISE-R \\
2017.63 &  8.22$\pm$0.01 &  7.62$\pm$0.01 &                & NEOWISE-R \\
2018.17 &  8.33$\pm$0.02 &  7.76$\pm$0.01 &                & NEOWISE-R \\
2018.62 &  8.29$\pm$0.01 &  7.72$\pm$0.02 &                & NEOWISE-R \\
\hline
\end{tabular}
\end{table}

\begin{figure}
\centering
\includegraphics[width=8.5cm]{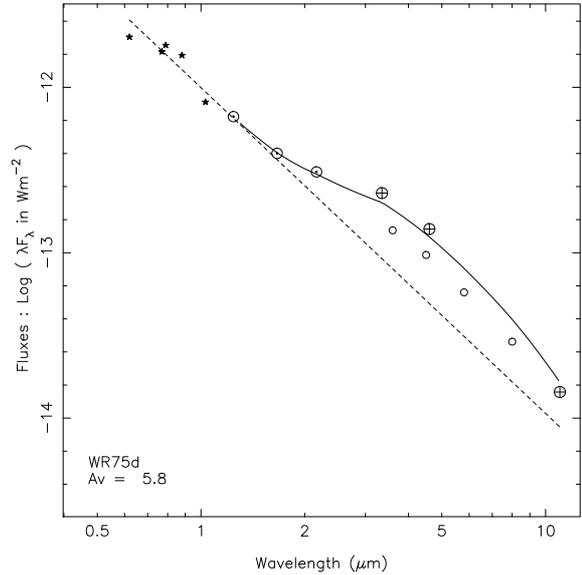}
\caption{SEDs of WR\,75d based on $W1$, $W2$ and $W3$ observed in 2010.67 in 
the 3-Band Cryo Survey (marked $\oplus$) and $JHK_s$ from 2MASS ($\odot$), 
with a stellar wind fitted to $r$,
$i, Z, Y$ ($\star$) and $J$ to get the reddening. Also plotted ($\circ$), 
but not used in the model fit, are fluxes observed in 2004.68 in the 
GLIMPSE survey, when the dust emission was evidently less.}
\label{F75d}
\end{figure}

This is another WC9 star identified by \citet{HBD}, but it was not considered 
to be a dust maker. The photometry is collected in Table~\ref{T75d}.
During 2010, in the six months between the All-Sky and 3-Band-Cryo 
observations, WR\,75d brightened significantly in $W1$ and $W2$ (but not $W3$) 
to its brightest observed in this progamme, and brighter than observed in the 
GLIMPSE Survey in 2004.

The SED using {\em WISE} 3-Band Cryo Survey data is plotted in Fig.\,\ref{F75d}; 
this sampled WR\,75d at its brightest and can be seen to lie above that from the 
GLIMPSE data. Although the latter (and the 2018 NEOWISE-R data) show WR\,75d 
at its faintest, the fluxes are still above the stellar wind level, suggesting 
that WR\,75d is not an episodic but a variable dust maker. No periodicity was 
found in the mid-IR photometry.

%----------------------------------------------------------------------------------
\subsection{WR\,75-11 = MDM 26} % ---------------------                    WR 75-11
%----------------------------------------------------------------------------------

\begin{table} 
\centering 
\caption{IR Photometric history of WR\,75-11.} 
\label{T75_11}
\begin{tabular}{lrrl}
\hline
  Date  & $W1$           &   $W2$          & Source\\
\hline
2004.25 &  9.30$\pm$0.05 &  8.73$\pm$0.08  & GLIMPSE \\ %e 0.05, 0.08
2010.17 &  9.50$\pm$0.02 &  8.91$\pm$0.02  & All-Sky \\
2010.67 &  9.26$\pm$0.03 &  8.65$\pm$0.02  & 3-Band Cryo \\ % W3 7.85 (0.14)
2012.75 &                &  8.74$\pm$0.03  & Deep GLIMPSE \\  % no [3.6]
2014.18 &  9.34$\pm$0.01 &  8.82$\pm$0.01  & NEOWISE-R \\
2014.68 &  9.40$\pm$0.03 &  8.90$\pm$0.01  & NEOWISE-R \\
2015.17 &  9.40$\pm$0.01 &  8.83$\pm$0.01  & NEOWISE-R \\
2015.66 &  9.50$\pm$0.02 &  9.00$\pm$0.01  & NEOWISE-R \\
2016.18 &  9.52$\pm$0.01 &  8.99$\pm$0.01  & NEOWISE-R \\
2016.64 &  9.53$\pm$0.02 &  9.02$\pm$0.02  & NEOWISE-R \\
2017.17 &  9.42$\pm$0.01 &  8.88$\pm$0.01  & NEOWISE-R \\
2017.63 &  9.41$\pm$0.01 &  8.88$\pm$0.01  & NEOWISE-R \\
2018.17 &  9.41$\pm$0.01 &  8.89$\pm$0.01  & NEOWISE-R \\
2018.62 &  9.43$\pm$0.01 &  8.91$\pm$0.01  & NEOWISE-R \\
\hline
\end{tabular}
\end{table}

\begin{figure}
\centering
\includegraphics[width=8.5cm]{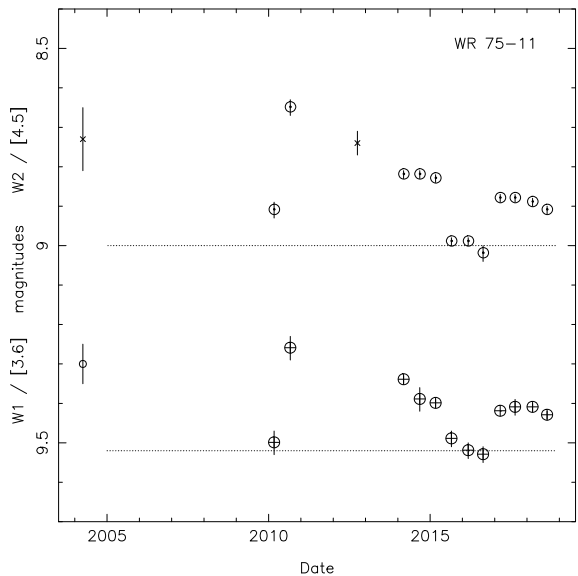}
\caption{Synoptic photometry of WR\,75-11 using $W1$ ($\oplus$), $W2$ ($\odot$), 
GLIMPSE [3.6] ($\circ$) and [4.5] ($\times$). Error bars are $\pm 1\sigma$. The 
dotted lines, from the 2016 data, indicate the dust-free continuum level.}
\label{Fsynoptic75_11}
\end{figure}

\citet{MDM11} identified WR\,75-11 (their MDM 26) as a WR star on the basis 
of its IR colours and classified its spectrum as `WC9d?'. They noted that the 
emission lines in their $H$- and $K$-band spectra observed on 2010 May 26 
were relatively weak, possibly as a result of dilution from thermal dust 
emission, but hesitated to classify the star as a dust maker because its IR 
colours were not characteristic of dust emission. 
The {\em WISE} data (Table \ref{T75_11}) help resolve this apparent 
discrepancy, showing that $W1$ and $W2$ brightened by about 0.25 mag.\ between 
2010.17 and 2010.67, consistent with a dust-formation episode occurring 
between these dates. 
After the 3-Band Cryo observation in 2010.67, the IR flux faded so that the 
VVV $K_s$ and Deep GLIMPSE [4.5] magnitudes in 2012 were close to their 
earlier $K_s$ and GLIMPSE values. The NEOWISE-R observations in 2014--2015 
showed continued fading in $W1$ and $W2$ (Fig.\,\ref{Fsynoptic75_11}). 
Even at maximum, the dust emission is marginal -- the ratio dust/wind at 
3.4 $\micron$ is only 0.2 and this soon faded, so WR\,75-11 should classed as 
an episodic dust maker showing stellar wind emission for some of the time, 
but with rather uneven dust formation episodes.

%-------------------------------------------------------------------------------
\subsection{WR\,77t = HBD 5}                   % ---------- $\Delta W1 = 0.39
%-------------------------------------------------------------------------------

\begin{table} 
\centering 
\caption{Mid-IR Photometric history of WR\,77t.} 
\label{T77t}
\begin{tabular}{lrrrl}
\hline
  Date  & $W1$  & $W2$  & $W3$  & Source\\
\hline
2004.68 &  7.04$\pm$0.04 &  6.51$\pm$0.04 &       & GLIMPSE \\ % [8.0] 5.92 (.02)
2010.18 &  7.09$\pm$0.01 &  6.44$\pm$0.01 & 6.03$\pm$0.02  & All-Sky \\ % W4 5.40
2010.67 &  7.00$\pm$0.01 &  6.30$\pm$0.01 & 6.01$\pm$0.03  & 3-Band Cryo \\
2014.19 &  6.82$\pm$0.02 &  6.23$\pm$0.01 &       & NEOWISE-R \\
2014.68 &  6.91$\pm$0.02 &  6.36$\pm$0.02 &       & NEOWISE-R \\
2015.18 &  6.96$\pm$0.03 &  6.41$\pm$0.02 &       & NEOWISE-R \\
2015.67 &  6.71$\pm$0.03 &  6.22$\pm$0.03 &       & NEOWISE-R \\
2016.18 &  7.14$\pm$0.03 &  6.56$\pm$0.03 &       & NEOWISE-R \\
2016.65 &  6.65$\pm$0.02 &  6.18$\pm$0.02 &       & NEOWISE-R \\
2017.18 &  6.71$\pm$0.03 &  6.23$\pm$0.02 &       & NEOWISE-R \\
2017.64 &  7.12$\pm$0.04 &  6.58$\pm$0.02 &       & NEOWISE-R \\
2018.18 &  6.81$\pm$0.02 &  6.34$\pm$0.03 &       & NEOWISE-R \\
2018.62 &  7.10$\pm$0.04 &  6.55$\pm$0.02 &       & NEOWISE-R \\
\hline
\end{tabular}
\end{table}

\begin{figure}
\centering
\includegraphics[angle=270,width=8.5cm]{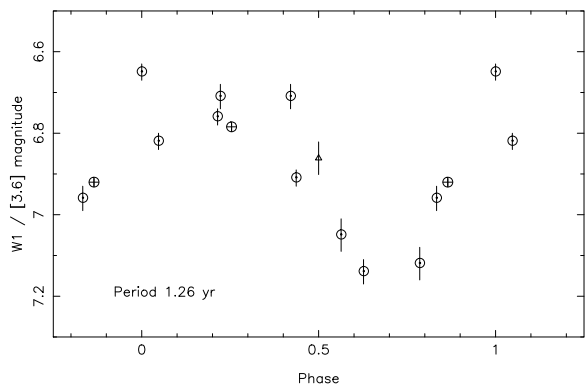}
\caption{$W1$ photometry of WR\,77t from the All-Sky and 3 Band Cryo ($\oplus$),
NEOWISE-R ($\odot$) and GLIMPSE ($\triangle$) surveys. The cryogenic data have 
been offset to match the NEOWISE-R data. Error bars are $\pm 1\sigma$.}
\label{F77t}
\end{figure}

This is another WC9 star identified by \cite{HBD}, who considered it to be 
a dust maker. Unfortunately, WR\,77t is brighter than NEOWISE-R saturation 
limits, which may account for some of the range seen in its NEOWISE-R 
photometry (Table~\ref{T77t}) and the differences from the cryogenic data. 
The dispersions in $W1$ and $W2$ for stars having similar brightness to 
WR\,77t ($\sigma$ 0.037 and 0.031, Section~\ref{SData}) suggest that the 
offsets between the NEOWISE-R and cryogenic magnitudes may be similar for 
each observation of any particular star of that brightness. Assuming this 
applies to WR\,77t, the NEOWISE-R data were searched for a period, yielding 
1.26~y. The NEOWISE $W1$ photometry phased to this period, together with the 
cryogenic data offset to match as derived from \citet[figure 6]{NEOWISER} 
are shown in Fig.\,\ref{F77t}. The discordance of some of the data suggests 
that the assumption of constant offset may be wrong, but it seems safe to 
conclude that WR\,77t is a variable on a time scale of 1--2 y. It is far 
too bright for a light curve from VVV $K_s$, but it is bright enough for a 
dedicated IR photometric study.

%------------------------------------------------------------------------------
\subsection{WR\,91-1 = SMG09 1222\_15}         % ---------------------- WR 91-1
%------------------------------------------------------------------------------

\begin{table} 
\centering 
\caption{Mid-IR Photometric history of WR\,91-1.} 
\label{T91_1}
\begin{tabular}{lrrl}
\hline
  Date  & $W1$  & $W2$  & Source\\
\hline
2005.72 &  9.34$\pm$0.04 &  8.70$\pm$0.06 & GLIMPSE II \\ % [5.8] 8.37, [8.0] 8.05
2006.32 &  9.16$\pm$0.05 &  8.72$\pm$0.07 & GLIMPSE II \\ % [5.8] 8.37, [8.0] 8.10
2010.20 &  9.57$\pm$0.02 &  8.65$\pm$0.02 & All Sky \\    % W3 = 8.32
2010.69 &  9.57$\pm$0.02 &  8.66$\pm$0.02 & 3-Band Cryo \\
2014.20 &  8.91$\pm$0.01 &  7.66$\pm$0.01 & NEOWISE-R \\
2014.70 &  9.15$\pm$0.01 &  7.90$\pm$0.01 & NEOWISE-R \\
2015.19 &  9.35$\pm$0.01 &  8.20$\pm$0.01 & NEOWISE-R \\
2015.68 &  9.44$\pm$0.03 &  8.35$\pm$0.01 & NEOWISE-R \\
2016.20 &  9.52$\pm$0.01 &  8.52$\pm$0.01 & NEOWISE-R \\
2016.66 &  9.56$\pm$0.01 &  8.57$\pm$0.01 & NEOWISE-R \\
2017.20 &  9.59$\pm$0.01 &  8.61$\pm$0.01 & NEOWISE-R \\
2017.65 &  9.61$\pm$0.01 &  8.68$\pm$0.01 & NEOWISE-R \\
2018.20 &  9.60$\pm$0.01 &  8.67$\pm$0.01 & NEOWISE-R \\
2018.65 &  9.61$\pm$0.01 &  8.73$\pm$0.01 & NEOWISE-R \\
\hline
\end{tabular}
\end{table}

\begin{figure}
\centering
\includegraphics[width=8.5cm]{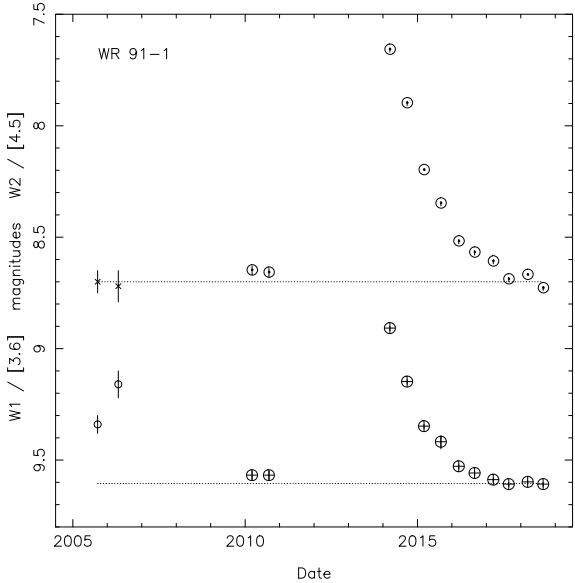}
\caption{Synoptic photometry of WR\,91-1 using $W1$ ($\oplus$), $W2$ ($\odot$), 
GLIMPSE [3.6] ($\circ$) and [4.5] ($\times$). Error bars are $\pm 1\sigma$. 
The dotted lines mark the likely continuum levels.}
\label{Fsynoptic91_1}
\end{figure}

This was identified as a WR star by \citet{SMG09}, who designated it 1222\_15 
and classified it as WC8 from its $K$-band spectrum. From a $J$-band spectrum, 
\citet{RC15} re-classified it as WC7. The only published photometry at shorter 
wavelengths is $Y$ from the VVV but it was possible to measure $Z \simeq 22.1$ 
in a 2-arcsec aperture on the $Z$ image retrieved from the VVV archive and to 
use that, together with $Y$ and $J$, to determine the reddening: $A_V$ = 27. 

The first NEOWISE-R observations in 2014.20 found WR\,91-1 to be significantly 
brighter than in the All-Sky and 3-Band Cryo surveys in 2010 (Table~\ref{T91_1}, 
Fig.\,\ref{Fsynoptic91_1}) 
and redder $(W1-W2) = 1.25$ compared with $(W1-W2) = 0.62$ previously. 
Subsequent NEOWISE-R observations (2014.70 -- 2018.65) showed steady fading 
towards the levels seen in the All-Sky and Post-Cryo surveys, $W2$ fading more 
slowly because the newly formed dust was cooling -- as seen in other 
episodic dust makers, cf. light curves of WR\,140 \citet[fig.\ 1]{W140d}.
Evidently, that there was a dust-formation episode some time between 
2010.69 and 2014.20. The 4.4-y duration of the mid-IR fading is too long for 
there to have been a dust formation and fading episode in the 3.9~y between 
the second GLIMPSE and first {\em WISE} observations, so the fading from any 
previous episode must have been complete by 2005, implying a period of at 
least 13~y.\ if such events are periodic.

%------------------------------------------------------------------------------
\subsection{WR\,122-14 = KSF14 1553-15DF}                 %  $\Delta W1$ = 0.17
\label{S122-14}
%------------------------------------------------------------------------------

\begin{table} 
\centering 
\caption{mid-IR Photometric history of WR\,122-14.} 
\label{T122_14}
\begin{tabular}{lrrl}
\hline
  Date  & $W1$  & $W2$  & Source\\
\hline
2004.31 &  9.54$\pm$0.04 &  8.78$\pm$0.03 & GLIMPSE \\
2010.25 &  9.73$\pm$0.03 &  8.66$\pm$0.02 & All-Sky \\
2010.75 &  9.62$\pm$0.03 &  8.51$\pm$0.03 & Post-Cryo \\
2012.44 &  9.34$\pm$0.04 &  8.54$\pm$0.03 & Deep GLIMPSE \\
2014.76 &  9.69$\pm$0.02 &  8.63$\pm$0.02 & NEOWISE-R \\
2015.26 &  9.77$\pm$0.02 &  8.72$\pm$0.02 & NEOWISE-R \\
2015.75 &  9.86$\pm$0.02 &  8.79$\pm$0.02 & NEOWISE-R \\
2016.26 &  9.87$\pm$0.02 &  8.81$\pm$0.02 & NEOWISE-R \\
2016.73 &  9.87$\pm$0.02 &  8.79$\pm$0.02 & NEOWISE-R \\
2017.26 &  9.72$\pm$0.01 &  8.68$\pm$0.01 & NEOWISE-R \\
2017.72 &  9.83$\pm$0.01 &  8.75$\pm$0.01 & NEOWISE-R \\
2018.26 &  9.57$\pm$0.01 &  8.50$\pm$0.01 & NEOWISE-R \\
2018.70 &  9.67$\pm$0.01 &  8.59$\pm$0.01 & NEOWISE-R \\
\hline
\end{tabular}
\end{table}

\begin{figure}                           % -----         WR 122-14 light curves
\centering
\includegraphics[width=8.5cm]{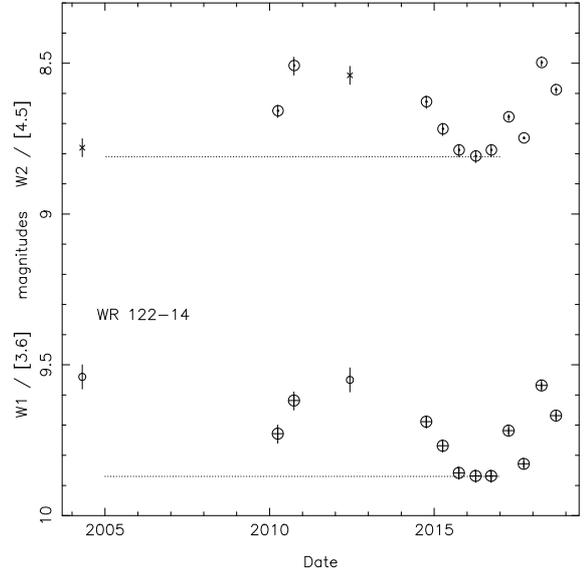}
\caption{Synoptic light curves of WR\,122-14 in $W1$ ($\oplus$), [3.6] ($\circ$), 
$W2$ ($\odot$) and [4.5] ($\times$). Error bars are $\pm$ 1 $\sigma$). The 
dotted lines suggest the continuum level in the high reddening case (see text).}
\label{Fsynoptic122_14}
\end{figure}

This was identified as a WR star by \citet{KSF14}, who classified it WC8. 
It brightened in $W1$ and $W2$ between the All-Sky and Post-Cryo survey 
observations (Table \ref{T122_14}, Fig.\,\ref{Fsynoptic122_14}) and brightened 
further to a broad maximum at the time of Deep GLIMPSE observations, and then 
faded to a constant level in 2015.75--2016.73. 
It is very heavily reddened: there appears to be no photometry shortward 
of 1 micron or even in $J$. From the $J$-band image in the UKIDSS Data 
Archive, it was possible to measure $J \simeq 19.7\pm0.3$. 
The SED can be fitted by a stellar wind reddened by $A_V$ = 48 using this $J$ 
and the UKIDSS $HK_s$, providing an upper limit to the reddening. 
This SED fits the NEOWISE-R data at minimum, supporting the suggestion that 
WR\,122-14 is an episodic dust maker,  but it is also possible that the 
reddening is less than estimated from $JHK_s$, so that dust emission was 
occurring in 2015--16 as well.

%------------------------------------------------------------------------------
\subsection{The previously known dust-maker WR\,125} 
%------------------------------------------------------------------------------

\begin{table} 
\centering 
\caption{{\em WISE} photometric history of WR\,125.} 
\label{T125}
\begin{tabular}{lccl}
\hline
  Date  &      $W1$      &      $W2$      & Survey\\
\hline
2010.29	&  7.79$\pm$0.02 &  7.40$\pm$0.02 & All-Sky \\  % W4 = 3.07
2010.79	&  7.77$\pm$0.02 &  7.29$\pm$0.02 & Post Cryo \\  % checked
2014.80	&  7.77$\pm$0.01 &  7.43$\pm$0.01 & NEOWISE-R \\
2015.30	&  7.78$\pm$0.01 &  7.45$\pm$0.01 & NEOWISE-R \\
2015.78	&  7.75$\pm$0.01 &  7.42$\pm$0.01 & NEOWISE-R \\
2016.29 &  7.76$\pm$0.01 &  7.45$\pm$0.01 & NEOWISE-R \\
2016.76 &  7.76$\pm$0.01 &  7.43$\pm$0.01 & NEOWISE-R \\
2017.29 &  7.79$\pm$0.01 &  7.44$\pm$0.01 & NEOWISE-R \\
2017.76 &  7.73$\pm$0.01 &  7.42$\pm$0.01 & NEOWISE-R \\
2018.30 &  7.69$\pm$0.01 &  7.36$\pm$0.01 & NEOWISE-R \\
2018.74 &  7.55$\pm$0.02 &  7.06$\pm$0.02 & NEOWISE-R \\
\hline
\end{tabular}
\end{table}

Because WR\,125 resembled WR\,140 in terms of its radio and X-ray emission 
and unusual breadth of its emission lines for its WC7 spectral type, it was 
monitored in the IR to search for evidence of dust emission, leading to 
discovery of a dust-formation episode starting in 1990 \citep{W125}. 
Further observations \citep{W125II} showed the IR flux reached a maximum in 
1992--93 and also found absorption lines in its spectrum, supporting its 
interpretation as a colliding wind binary. The IR emission faded and, by 
the date (1997 June 16) of the 2MASS observation, the 2-micron flux 
($K_s$ = 8.21) had faded close to the 1981--89 pre-outburst mean ($K$ = 8.25).

The 2014--17 NEOWISE-R observations (Table~\ref{T125}) show WR\,125 at 
constant levels, $W1$ = 7.76 ($\sigma$ 0.02) and $W2$ = 7.43 ($\sigma$ 0.01),  
close to the 1988--89 pre-outburst means $L^{\prime}$ = 7.75 and $M$ = 7.32.
The 2018 observations, however, show slow brightening, to $W1$ = 7.55$\pm$0.01
and $W2$ = 7.06$\pm$0.01 in 2018.74, 0.18 and 0.32 mag.\ above the wind level 
and redder than it, indicating the beginning of another dust formation event. 
Comparison with the sparse $L^{\prime}$ and $M$ photometry in 1990--91 
\citep{W125} suggests an interval of about 28.3~y between the episodes. 
Further observations are needed to define the dust formation and confirm its 
periodicity. 
Confirmation that WR\,125 is indeed brightening in the IR has been provided 
by recent ground-based photometry by Shenavrin (2019, private communication).
\citet*{WR125Xray} found no variation in the X-ray emission from four 
observations in 2016--17, giving no suggestion of approach to periastron 
passage, but re-observation now would be valuable.

%------------------------------------------------------------------------------
\subsection{WR\,125-1 = HDM 15}                  % ------------------- WR 125-1
%------------------------------------------------------------------------------

\begin{table} 
\centering 
\caption{IR Photometric history of WR\,125-1.} 
\label{T125-1}
\begin{tabular}{lcccl}
\hline
  Date  &     $K_s$      &      $W1$      &      $W2$      & Source\\
\hline
1997.46 &  9.07$\pm$0.02 &                &                & 2MASS \\
2004.78 &                &  8.17$\pm$0.02 &  7.78$\pm$0.03 & GLIMPSE \\ % also [5.8] = 7.43, [8.0] = 7.03
2006.52 &  9.32$\pm$0.00 &                &                & UKIDSS \\ % multiframeID 4547332 July 8 also J 10.39
2010.30	&                &  8.23$\pm$0.02 &  7.58$\pm$0.02 & All-Sky \\  % $W3$ = 6.36
2010.80	&                &  8.21$\pm$0.01 &  7.50$\pm$0.02 & Post-Cryo \\
2014.31	&                &  8.49$\pm$0.01 &  8.00$\pm$0.01 & NEOWISE-R \\
2014.80	&                &  8.46$\pm$0.01 &  7.99$\pm$0.01 & NEOWISE-R \\
2015.30	&                &  8.49$\pm$0.01 &  8.02$\pm$0.01 & NEOWISE-R \\
2015.79	&                &  8.54$\pm$0.01 &  8.07$\pm$0.01 & NEOWISE-R \\
2016.30 &                &  8.54$\pm$0.01 &  8.08$\pm$0.01 & NEOWISE-R \\
2016.77 &                &  8.54$\pm$0.01 &  8.11$\pm$0.01 & NEOWISE-R \\
2017.30 &                &  8.57$\pm$0.01 &  8.14$\pm$0.01 & NEOWISE-R \\
2017.76 &                &  8.54$\pm$0.01 &  8.12$\pm$0.02 & NEOWISE-R \\
2018.30 &                &  8.58$\pm$0.01 &  8.16$\pm$0.01 & NEOWISE-R \\
2018.75 &                &  8.56$\pm$0.01 &  8.16$\pm$0.01 & NEOWISE-R \\

\hline
\end{tabular}
\end{table}

\begin{figure}                                              % ---- WR 125-1 SED
\centering
\includegraphics[width=8.5cm]{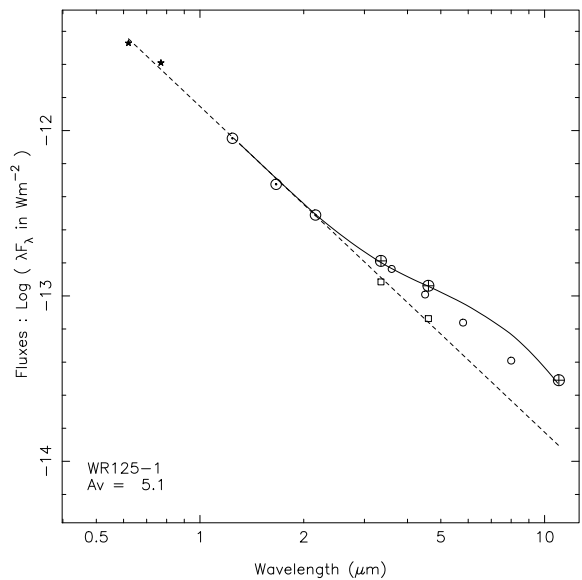}
\caption{SED of WR\,125-1 based on the 2010 AllWISE $W1$, $W2$ and $W3$ 
($\oplus$) with a stellar wind fitted to $r$, $i$ and $J$. Also plotted, 
but not used in the model fit, are fluxes ($\circ$) from the GLIMPSE 
photometry in 2004, showing less dust emission, and those from the 
2018 NEOWISE-R observations ($\Box$), which fit the wind SED and 
indicate that the dust emission had faded by then.
}
\label{F125-1}
\end{figure}

WR\,125-1 was identified as a WR star by \citet{HDM} from its IR colours, 
designated HDM15 and classified WC8 from the C\,{\sc iv}/C\,{\sc iii} 
line ratios in the $Z$- and $J$ bands, confirmed by comparison of its 
$J$-band spectrum with that of the WC8 spectral standard WR\,135. 
Examination of this comparison (their fig 4) suggests that the emission 
lines in HDM15 are about half as strong as those in WR\,135. 
The same weakness is seen in other wavelength regions: the equivalent 
widths (EWs) of the 0.971-$\micron$ C\,{\sc iii} and 0.990-$\micron$ 
C\,{\sc iv} features are about half those of the corresponding features 
in WR\,135 measured by \citet{HowarthIR}, while that of the 2.076-$\micron$ 
C\,{\sc iv} feature is also about half that of the corresponding feature 
in the WR\,135 spectrum observed by \citet{Wvinf}. These differences 
suggest that the WC8 spectrum of HDM15 is diluted by another continuum 
source having about the same luminosity as the WC8 star; the extension 
of the dilution to wavelengths as short as 0.971~$\micron$ argues strongly 
against dilution by heated dust emission and suggests the presence of a 
line-of-sight or binary companion to the WC8 star in WR\,125-1.
The UKIDSS $J$ image was searched for a line-of-sight companion that 
could have contaminated the spectrum but none was found; of neighbours 
within 5~arc sec, the brightest is almost 4 mag fainter in $J$. 0.02

It is therefore probable that the WC8 star has a luminous companion
which should be searched for spectroscopically.

The NEOWISE-R $W1$ and $W2$ are significantly fainter than those observed 
in the All-Sky and Post-Cryo surveys (Table~\ref{T125-1}) and indicate 
that the relatively modest dust emission observed earlier had faded so as 
to be unobservable relative to the wind (Fig.\,\ref{F125-1}). The $K_s$ 
photometry suggests that the maximum had occurred prior to the 2MASS 
observation in 1997, but the gap in coverage between that and subsequent 
observations is too long to rule out a shorter time-scale.
The GLIMPSE observations in 2004.78 show WR\,125-1 at an intermediate level.
Evidently WR\,125-1 is another episodic dust maker.

%------------------------------------------------------------------------------
\subsection{HD 36402 = BAT99--38 = Br 31}  
%------------------------------------------------------------------------------

\begin{table} 
\centering 
\caption{{\em WISE} photometric history of HD 36402.} 
\label{T36402}
\begin{tabular}{lccl}
\hline
  Date  & $W1$  & $W2$  & Survey\\
\hline
2010.41	&  9.44$\pm$0.02 &  8.57$\pm$0.02 & All-Sky \\  
2010.91	&  9.25$\pm$0.01 &  8.38$\pm$0.01 & Post-Cryo \\
2014.42 &  9.98$\pm$0.01 &  9.11$\pm$0.01 & NEOWISE-R \\
2014.92	&  9.78$\pm$0.01 &  8.93$\pm$0.01 & NEOWISE-R \\
2015.41	&  9.53$\pm$0.01 &  8.69$\pm$0.01 & NEOWISE-R \\
2015.90	&  9.32$\pm$0.01 &  8.48$\pm$0.01 & NEOWISE-R \\
2016.41 &  9.29$\pm$0.01 &  8.43$\pm$0.01 & NEOWISE-R \\
2016.89 &  9.48$\pm$0.01 &  8.60$\pm$0.01 & NEOWISE-R \\
2017.41 &  9.75$\pm$0.01 &  8.86$\pm$0.01 & NEOWISE-R \\
2017.89 & 10.05$\pm$0.01 &  9.20$\pm$0.02 & NEOWISE-R \\
2018.41 & 10.17$\pm$0.01 &  9.33$\pm$0.01 & NEOWISE-R \\
2018.88 & 10.18$\pm$0.01 &  9.32$\pm$0.01 & NEOWISE-R \\
\hline
\end{tabular}
\end{table}

Spectroscopy of HD~36402 in the LMC suggests that it is a triple system. 
From the WR emission lines, \citet{Orb36402}, derived an 3.03-d orbit which 
was not shared by the absorption lines, suggesting that the O8 supergaint in 
which the latter formed did not participate in the 3-d orbit, but was in a 
longer period orbit about the inner WC4+O? binary. Variations in the dust 
emission from HD 36402 based on IR data from a variety of sources was 
reported by \citet{W36402}. From the brightening in 2004--05 and 2009--10, 
they derived a period near 4.7~y., which they associated with the outer 
orbit of the O8 supergiant. They found no evidence for variation related to 
the 3.03-d. orbit in the relatively long visits (21 and 17~d. respectively) 
in the {\em WISE} All-Sky and Post-Cryo surveys. 

\begin{figure}                                                  % light curves
\centering
\includegraphics[angle=270,width=8.5cm]{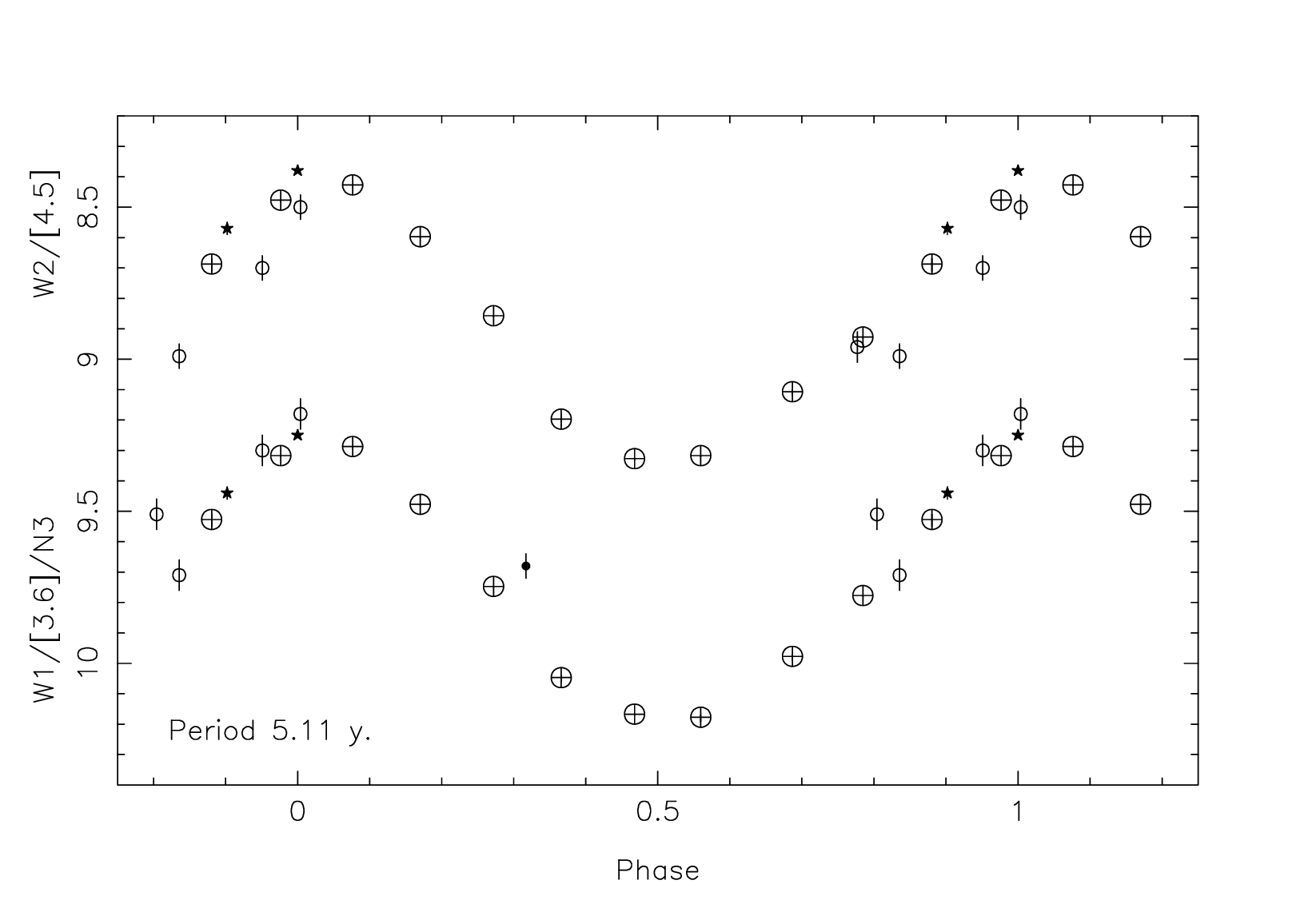}
\caption{Phased mid-IR light curves of HD 36402 using {\em WISE} All-Sky and 
Post-Cryo ($\star$) and NEOWISE-R ($\oplus$) data, together with IRAC ($\circ$), 
and {\em AKARI} $N3$ ($\bullet$); see \citet{W36402} for sources of the latter.
Error bars are $\pm$ 1 $\sigma$.}
\label{F36402}
\end{figure}

The NEOWISE-R data (Table~\ref{T36402}) cover practically a whole period, with 
a maximum in early 2016, later than expected from the 4.7-y.\ period, followed 
by fading. Re-determination of the period including the NEOWISE-R data suggests 
a period near 5.11~y.; the {\em WISE} data alone give a period of 5.2~y. 
A phased light curve is given in Fig.\,\ref{F36402}, where zero phase is set 
to the epoch of the {\em WISE} Post-Cryo observation, close to maximum. 
The variations are slow, taking $\sim$ 1.5 y. to rise to maximum 
and the same to fade to minimum.
The colour remains significantly greater than that expected of the stellar wind, 
$W1$--$W2$ = 0.17, indicating continuous dust emission. It barely changes, from 
an average of $W1$--$W2$ = 0.85$\pm$0.01 in the three visits (2014.4--2015.4) 
before maximum, when we would expect a higher fraction of hotter, newly-formed 
dust, to 0.87$\pm$0.01 in the three visits (2016.9-2017.9) after maximum, when 
one would expect a higher fraction of cooler dust while the emission fades. 

Spectroscopy of the O8 supergiant over a period of years is needed to test 
whether it is a member of a CWB and, if the orbit was elliptical, the relation 
between its periastron passage and the maximum in dust formation.

%------------------------------------------------------------------------------
\subsection{HD 38030 = BAT99--84 = Br 68}         
%------------------------------------------------------------------------------

\begin{table} 
\centering 
\caption{Mid-IR photometric history of HD 38030.} 
\label{T38030}
\begin{tabular}{lccl}
\hline
  Date  & $W1$  & $W2$  & Survey\\
\hline
2005 55 & 12.80$\pm$0.04 & 12.53$\pm$0.05 &  SAGE Epoch 1 \\
2005.83 & 12.81$\pm$0.06 & 12.60$\pm$0.07 &  SAGE Epoch 2 \\
2010.31 & 12.68$\pm$0.02 & 12.48$\pm$0.02 &  All-Sky \\
2010.80 & 12.72$\pm$0.02 & 12.48$\pm$0.02 &  Post-Cryo \\
2014.32 & 12.69$\pm$0.01 & 12.53$\pm$0.01 & NEOWISE-R \\
2014.82 & 12.70$\pm$0.01 & 12.51$\pm$0.01 & NEOWISE-R \\  
2015.31 & 12.71$\pm$0.01 & 12.52$\pm$0.01 & NEOWISE-R \\  
2015.80 & 12.70$\pm$0.01 & 12.52$\pm$0.01 & NEOWISE-R \\  
2016.31 & 12.71$\pm$0.01 & 12.54$\pm$0.01 & NEOWISE-R \\  
2016.80 & 12.70$\pm$0.01 & 12.54$\pm$0.01 & NEOWISE-R \\  
2017.31 & 12.70$\pm$0.01 & 12.53$\pm$0.01 & NEOWISE-R \\  
2017.77 & 12.71$\pm$0.01 & 12.54$\pm$0.01 & NEOWISE-R \\   
2018.31 & 12.00$\pm$0.01 & 11.26$\pm$0.01 & NEOWISE-R \\   
2018.43 & 11.87$\pm$0.02 & 11.05$\pm$0.02 & NEOWISE-R \\   
2018.76 & 11.52$\pm$0.01 & 10.62$\pm$0.01 & NEOWISE-R \\   
\hline
\end{tabular}
\end{table}

\begin{figure}                                              % ---- HD 38030 SED
\centering
\includegraphics[width=8.5cm]{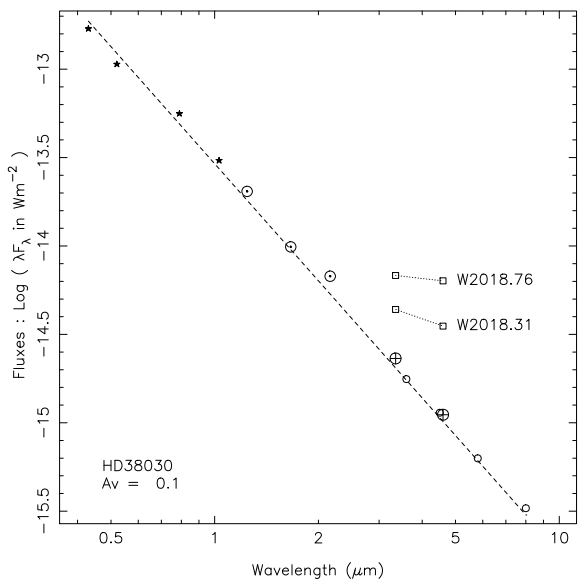}
\caption{SED of HD 38030 based on the AllWISE $W1$ and $W2$ ($\oplus$), SAGE 
[3.6] to [8.0] ($\circ$), 2MASS $JHK_s$ ($\odot$), with a stellar wind fitted to 
$b$, $v$ \citep{SmithNB}, $i$ \citep{DENIS} and $Y$ \citep{VMC1} representing 
HD~38030 in quiescence. Plotted above this SED are fluxes from the NEOWISE-R 
$W1$ and $W2$ observed in 2018.31 and 2018.76 ($\Box$) showing the change in 
level and slope of this portion of the mid-IR SED.}
\label{F38030}
\end{figure}

During 2010--2017, the {\em WISE} photometry of HD 38030 (Table~\ref{T38030}) 
showed no variation but, in 2018, the flux was observed to be rising sharply, 
at rates exceeding 1~mag~y$^{-1}$. 
The first 2018 visit was sufficiently protracted that it could be split into 
two and mean magnitudes calculated for the 
separated segments, giving three observations for 2018. A preliminary report, 
including light curves, is given by \citet{W38030}. The $W1-W2$ colour 
was also significantly greater than that of the wind emission. This is shown 
in Fig.\,\ref{F38030}, where the fluxes from $W1$ and $W2$ from the 2018.31 
and 2018.76 observations can be compared with the wind SED. Unfortunately, 
there is no contemporaneous IR photometry at other wavelengths to define the 
SED but the $W1-W2$ colour temperature, $\sim$ 830~K from $W1$ and $W2$, is 
too low for any photospheric emission and points to heated circumstellar dust. 

This appears to be the first recorded such event from HD 38030; before the 
first {\em WISE} observations, it was observed in 2005 in the SAGE \citep{SAGEphot} 
survey, having [3.6] and [4.5] magnitudes consistent with the {\em WISE} data. 
In the near-IR, it had comparable $K_s$ magnitudes in the 2MASS 6X Point Source 
Working Database \citep{LMC6X} in 2000--01, the IRSF \citep{IRSF} in 2003 and 
the Vista Magellanic Clouds Survey \citep{VMC1} in 2010. Taken together, these 
indicate quiescence for at least 17~y. If dust-formation episodes are recurrent, 
the previous one must have occurred some time before 1998, to allow time for 
the dust to cool and the emission to fade, implying a period in excess of 20~y. 

There is currently no other evidence that HD 38030 is a CWB. \citet{ChandraLMC} 
did not detect X-ray emission from HD 38030 in their survey of LMC WR stars using 
{\em Chandra} ACIS.
From the lack of variability in RVs observed in 1984 and 1993, \citet*{Bartzakos} 
deduced that HD 38030 was almost certainly a single star and that the absorption 
lines in its spectrum might arise in a visual OB companion but, given the long 
period and likely eccentric orbit, the time-span of RV variations might be quite 
short.

% 2MASS \citet{2MASS} 2000.10 K 12.96 (.04)
% LMC 6X \citet{LMC6X} 2000.94  K 12.86 (03) 2MASS 6X Point Source Working Database
% LMC 6X \citet{LMC6X} 2001.10  K 12.88 (03)
% IRSF \citet{IRSF} 2003.86 K = 12.90
% VMC \citet{VMC1}  2010.02 K = 12.90

%------------------------------------------------------------------------------
\section{Discussion}                                         % ----------------
%------------------------------------------------------------------------------

\subsection{Incidence of dust emission}

On the basis of their SEDs, 17 Galactic WR stars not previously considered to 
be dust makers were found here to show dust emission: ten apparently constant 
(Table~\ref{Tnewdust}) and seven variable (Table~\ref{Tnewvar}), to which one 
can add one of the stars (WR\,75-11) originally classified as a possible dust 
maker. This represents more than one-tenth of the present sample, 
despite the fact that the NEOWISE-R survey is not the best data set for a 
census of dust formation by WR stars owing to selection effects. 
First, the saturation limits exclude the IR brighter stars, not only almost 
all of the previously known WC8--9 dust makers, but also recently identified 
WC9 stars such as WR\,111-12 \citep{MM14} which, in the process of inspecting 
data for the present study, was found to have a bright dust-emission SED. 
Secondly, the confusion limits excluded more stars in crowded regions, 
introducing a bias against those in the inner Galaxy, where dust-making 
WR stars are more common \citep{RC15}. Therefore, the present study calls for 
a re-assessment of the incidence of dust formation by WC stars, 

The incidence of dust emission amongst the population of Galactic WC stars as 
a whole is likely to be underestimated owing to a further layer of selection: 
many of the more recently identified WR stars in the Catalogue were 
discovered on the basis of WR line emission in their 2-$\micron$ spectra 
e.g. \citep{Homeier, SMG09, SFZ12, KSF14}, which will be diluted if there is 
dust emission in that wavelength region.  This introduces a bias against 
WR stars having the strongest dust emission, so that the true fraction of 
dust-emitters amongst WC stars in the Galaxy is therefore likely to be 
significantly underestimated. A search for WR stars based on the $J$-band 
spectrum, which includes strong emission lines and is less affected by dust 
emission, could obviate this problem.

The identification of a second WR dust maker in the LMC confirms that this 
process can occur in stars formed in lower metallicity environments and raises 
the question of what is the lowest metallicity environment in which WR stars 
can have strong enough winds to form dust in CWBs.

\begin{table*}
\centering
\caption{Properties of WR stars showing variation in their dust emission, including 
orbital information or indications of possible binarity. Episodic dust makers are 
flagged `Ep' and peristent variable dust makers `V', along with dates of maxima, 
more relevant for the longer period systems. The amplitudes $\Delta L$ are those in 
$W1$ for the variables found in this study and in $L^{\prime}$ for the previously known 
variables. The periods are of the dust emission.}
\begin{threeparttable}
\begin{tabular}{llclllll}
\hline
Star      & \multicolumn{3}{c}{--- Stellar system properties ---} & \multicolumn{4}{c}{--- Dust emission phenomena and properties ---}   \\ 
          &  spectrum    & Binarity status         & Refs & Var. type, dates of IR maxima & $\Delta L$ & P(y)    & Refs \\
\hline
WR\,19    &  WC5 + O9    & $e$ = 0.8 ($P$ from IR) & 1, 2 & Ep., (1987), 1997, 2007, 2018     & $>$2.0 & 10.1          & 3    \\
WR\,46-7  &  WC5-7       &                         &      & V., 2010.57, 2015.08, 2016.55,    & 0.86   & 1.49          &      \\
WR\,47c   &  WC5         &                         &      & Ep., 2005--10 (broad maximum)     & 0.41   & $>$ 14        &      \\ 
WR\,48a   &  WC8 + WN8   &                         & 4    & V., 1979, 2011;      pinwheel     & 2.85   & 32.5          & 5, 6 \\
WR\,60-4  &  WC8         &                         &      & Ep., 2015.60, 2017.12             & 0.28   &  0.76         &      \\
WR\,65    &  WC9 + OB    &  absorption lines       & 7    & V., 1979-80                       & 0.41   & $\simeq$ 4.8  & 7    \\
WR\,70    &  WC9 + B0I   &  SB2                    & 8    & V., 1989, 1997, 2008, irregular   & 0.6    & 2.8 ?         & 9    \\
WR\,75aa  &  WC9d        &                         &      & V., 2010.66, 2015.66, 2018.61     & 0.17   & irreg.        &      \\
WR\,75d   &  WC9         &                         &      & V., 2010.67                       & 0.32   &               &      \\
WR\,75-11 &  WC9d?       &                         &      & Ep., 2010.67                      & 0.26   &               &      \\
WR\,77t   &  WC9d        &                         &      & V., 2016.65                       & 0.49   & 1.26          &      \\ 
WR\,91-1  &  WC7         &                         &      & Ep., 2014.20                      & 0.66   & $>$ 13        &      \\
WR\,98a   &  WC8-9       &                         &      & V., rotating pinwheel, P 1.54 y   & 0.92   & 1.54          & 10, 11 \\
WR\,112   &  WC8-9       & diluted emission lines  & 12   & V., pinwheel                      &        & 12.3          & 7    \\
WR\,122-14 & WC8         &                         &      & Ep., 2010.75, 2018.26             & 0.25   &               &      \\
WR\,125   &  WC7 + O9    & absorption lines        & 13   & Ep., 1992.7                       & 2.75   & $\sim$ 28.3   & 13   \\
WR\,125-1 &  WC8         & diluted emission lines  &      & Ep., 2010.80                      & 0.33   &               &      \\ 
WR\,137   &  WC7 + O9 	 & $P$ = 13.05~y, $e$ = 0.18 & 14 & Ep., 1984, 1997, 2010             & 1.59   & 13.05         & 15    \\
WR\,140   &  WC7 + O5    & $P$ = 7.93~y, $e$ = 0.8964 & 16 & Ep., 1977, 1985, 1993, 2001, 2009 & 2.57  & 7.94          & 17, 18 \\
HD 36402  &  WC4(+O)+O8I &  triple system          & 19   & V., c.1996.9, 2011, 2016          & 0.73   & 4.7           & 20   \\
HD 38030  &  WC4 + OB    &                         &      & Ep., 2018 or later                & 1.2    & $>$ 20        &    \\
\hline
\end{tabular}
\begin{tablenotes}
\item References: 
1 \citet{W19orb}; 2 \citet{CMB}; 3 \citet{Veen19}; 4 \citet{Zhekov48a}; 5 \citet{Marchenko48a112}; 6 \citet{W48a}; 
7 \citet{WPotsdam}; 8 \citet{Virpi70}; 9 \citet{W70}; 10 \citet{Monnier98a}; 11 \citet{WIAU212}; 12 \citet{CK76}; 
13 \citet{W125II}; 14 \citet{Orbit137}; 15 \citet{W137II};
16 \citet{Fahed140,Monnier140}; 17 \citet{W140,W140d}; 18 \citet{Taranova140}; 19 \citet{Orb36402}; 20 \citet{W36402}.
\end{tablenotes}
\end{threeparttable}
\label{Tprops}
\end{table*}

\subsection{Characterising the newly found dust variables}

The properties of the new dust variables are collected in Table~\ref{Tprops}, 
where they can be compared with the generally better known properties of the 
previously known systems, including those too bright for NEOWISE-R.
The Table does not include apparently constant dust makers which might be  
low-amplitude variable dust makers, as remarked in Section~\ref{SSEDs} above. 
The same possibility applies to the dust makers too bright for this study, 
such as those considered by \citet{WPotsdam}. 
The boundary between constant and variable in small data sets is hard to fix 
precisely.  In CWB dust formation paradigm, stars near this boundary may be 
members of binary stsems having near circular orbits.

Two of the previously known dust variables in Table~\ref{Tprops} call for comment. 
An extensive IR photometric study of WR\,70 (HD 137603) showed variation in dust 
emission on a range of time scales with a possible period near 2.82~y., but the 
variations were not strictly regular \citep{W70}. 
From the anti-correlation of the RVs of absorption and emission lines, the system 
is believed to a double-lined spectroscopic binary \citep{Virpi70}, but it still 
lacks an orbit. This deficiency needs to be met, not only to solve the system but 
also to help understand the variations in dust formation. This may, in turn, help 
understand apparently irregular variables having far fewer IR observations, like 
WR\,75aa. The second star, WR\,112 (CRL 2104), was considered to be variable on the 
basis of sequences of near-IR photometry in 1988--89 and 1997--2002 showing fading. 
The 12.3-y.\ suggested by their separation \citep{WPotsdam} was consistent with  
modelling of its dust pinwheel \citep{Marchenko112, Marchenko48a112}. Recently, 
however, new observations by \citet{Stagnant112} found that the dust structure 
about WR\,112 was not expanding, with no evidence for long term changes in the 
mid-IR flux, throwing the 12.3-y.-period into doubt and, indeed, opening up 
questions about WR dust formation which will have to be addressed elsewhere.
 
Given under the Stellar system properties in the Table are indications of 
binarity, ranging from full orbits to dilution of emission lines. These are 
sparse for the previously known systems and almost non-existent for the new 
variables found here. Filling these gaps is important if we are to understand the 
properties of dust making WR stars: although the determination of orbits for 
systems with long dust-emission periods is daunting, orbits for the brighter, 
shorter period systems are tractable.

It is apparent that most of the episodic variables (flagged `Ep') have earlier 
spectral subtype, with only four having subtypes WC8 or WC9, but this may be 
influenced by the following selection effect. The chance of observing stellar 
wind emission after dust emission has faded depends on the interplay of the 
fading time and the interval between dust-formation episodes. The rate at 
which the dust emission fades depends on the speed of the wind carrying away 
the dust, diluting the stellar flux heating it so that the dust cools. 
Stellar winds are generally faster for stars of earlier subtype, 
so there is a greater probability of observing dust-free emission from an 
episodic dust maker of earlier spectral subtype than one of later spectral 
subtype having a slower wind even if the periods are the same. An episodic dust 
maker with a slow  wind may be mistaken for a variable dust maker if there was 
not enough time for the dust emission to have faded to below wind level 
before the next dust formation episode began.
Consequently, the observational boundary between episodic and variable dust 
formation is uncertain; modelling of the evolution of the SED from IR 
phototmetry over a suitably wide range of wavelength should indicate if dust 
formation continues at a slower rate or ceases.

The duration of observable dust emission from an episodic dust maker also 
depends on the wavelength at which the emission is observed. 
This is very apparent in the multi-wavelength light curves of WR\,140 
\citep[fig. 1]{W140d}, which show the mid-IR flux taking a lot longer to 
fade than the near-IR data. This gives the {\em WISE} $W1$ and $W2$ bands 
an advantage in the search for episodic dust makers, 
but it is still possible that there exist in the data set undiscovered 
episodic dust makers having very long periods such that the sequence of 
{\em WISE} observations sampled the wind emission only.

\section{Conclusions}

Photometry from the NEOWISE-R Survey for a sample of 128 Galactic and 12 LMC 
WC-type WR stars selected to avoid source confusion or saturation has been 
searched for variability and evidence for circumstellar dust emission. 
Most of the stars were found to have dispersions in their $W1$ and $W2$ 
photometry of less than 5 per cent. Apart from three previously known 
variables and a wind eclipsing system found here, all the mid-IR variables 
are dust emitters. 
These include six episodic and four persistent variable dust makers in the 
Galaxy and one of each in the LMC. Eight of these stars were newly found 
to be dust makers in this study. 
Examination of the SEDs of the apparently non variable stars found a further 
ten new dust makers. This discovery of additional dust makers, and the 
selection effects which make this an underestimate, demonstrate the necessity 
for reinvestigation of the incidence of dust formation by WR stars.

The spectral subtypes of the new episodic dust makers have a wider range than 
those previously known, and include three WC8 and one WC9 star, while those of 
the variable persistent dust makers are mostly WC8--9 stars. The distinction 
between episodic and variable persistent dust makers favours identification of 
earlier subtype systems as episodic, so the distinction is not watertight. 
Also, three of the newly found constant dust makers have spectral subtypes 
earlier than WC8--9 -- the first such stars showing this property. The 
distribution of dust formation properties by spectral subtype is more nuanced 
than previously thought, but there is still a gradation of decreasing 
variability as one moves from earlier to later subtype.

Of the previously known episodic dust makers, the NEOWISE-R observations of 
WR\,19 captured its expected rise to maximum in 2017-18, while those of WR\,125 
the beginning of a new dust-formation episode, suggesting a period near 28.3~y.\ 
for the latter. This would be the longest period episodic dust maker known. 
The fragments of light curves of new dust makers found in the NEOWISE-R and 
other IR data suggest periods in excess of a decade for some and in excess 
of 20~y.\ for the newly found episodic dust maker HD 38030 in the LMC, 
calling for continued IR photometry of these systems for many years to come.

\section*{Acknowledgements}

This publication makes use of data products from the Near-Earth Object Wide-field 
Infrared Survey Explorer (NEOWISE-R), which is a project of the Jet Propulsion 
Laboratory/California Institute of Technology. NEOWISE-R is funded by the National 
Aeronautics and Space Administration. Data were retrieved from the NASA/ IPAC 
Infrared Science Archive, which is operated by the Jet Propulsion Laboratory, 
California Institute of Technology, the WFCAM Science Archive, VISTA Science 
Archive and OmegaCAM Science Archive, which are operated by the Wide Field 
Astronomy Unit of the Institute for Astronomy, University of Edinburgh, and 
the VizieR catalogue access tool, operated by the CDS, Strasbourg.
It is a pleasure to thank Victor Shenavrin for a timely observation of WR\,125 
and to acknowledge continued hospitality from the Institute for Astronomy.

\bibliographystyle{mnras}
\bibliography{neowise}                           % i.e. text/neoWISE/neowise.bib

% Don't change these lines
\bsp % typesetting comment
\label{lastpage}                   
\end{document}